\documentclass[twocolumn]{aastex63}

\usepackage{microtype} 
\usepackage{enumitem}
\usepackage{gensymb}
\usepackage{amsmath}
\usepackage{subfigure}

\newcommand{\flt}{\texttt{flt}}
\newcommand{\flc}{\texttt{flc}}
\newcommand{\skysub}{S_{\text{sub}}}
\newcommand{\skysubrms}{\sigma_{\text{sub}}}
\newcommand{\skyim}{S_{\text{chip}}}
\newcommand{\skyrmsim}{\sigma_{\text{chip}}}
\newcommand{\MJysr}{MJy sr$^{-1}$}
\newcommand{\nWmsr}{nW m$^{-2}$ sr$^{-1}$}

\newcommand{\new}[1]{#1}
\newcommand{\newnew}[1]{#1}

\defcitealias{carleton_2022}{SKYSURF-2}
\defcitealias{windhorst_2022}{SKYSURF-1}
\defcitealias{wfc3_ihb}{The WFC3 Instrument Handbook}
\defcitealias{acs_dhb}{The ACS Data Handbook}

\graphicspath{{./}{figures/}}

\begin{document}

\title{SKYSURF-4: Panchromatic HST All-Sky Surface-Brightness Measurement Methods and Results}

\correspondingauthor{Rosalia O'Brien}
\email{robrien5@asu.edu}

\author[0000-0003-3351-0878]{Rosalia O'Brien}
\author[0000-0001-6650-2853]{Timothy Carleton}
\author[0000-0001-8156-6281]{Rogier A. Windhorst}
\author[0000-0003-1268-5230]{Rolf A. Jansen}
\affiliation{School of Earth and Space Exploration, Arizona State University, Tempe, AZ 85287-1404, USA}

\author{Delondrae Carter}
\affiliation{School of Earth and Space Exploration, Arizona State University, Tempe, AZ 85287-1404, USA}

\author[0000-0001-9052-9837]{Scott Tompkins}
\affiliation{The University of Western Australia, M468, 35 Stirling Highway, Crawley, WA 6009, Australia}

\author[0000-0001-6990-7792]{Sarah Caddy}
\affiliation{Macquarie University, Sydney, NSW 2109, Australia}

\author[0000-0003-3329-1337]{Seth H. Cohen}
\affiliation{School of Earth and Space Exploration, Arizona State University, Tempe, AZ 85287-1404, USA}

\author{Haley Abate}
\affiliation{School of Earth and Space Exploration, Arizona State University, Tempe, AZ 85287-1404, USA}

\author[0000-0001-8403-8548]{Richard G. Arendt}
\affiliation{UMBC/CRESST2, NASA Goddard Space Flight Center, Greenbelt, MD 20771, USA}

\author[0000-0001-6265-0541]{Jessica Berkheimer}
\affiliation{School of Earth and Space Exploration, Arizona State University, Tempe, AZ 85287-1404, USA}

\author[0000-0002-0882-7702]{Annalisa Calamida}
\affiliation{Space Telescope Science Institute, 3700 San Martin Drive, Baltimore, MD 21210, USA}

\author{Stefano Casertano}
\affiliation{Space Telescope Science Institute, 3700 San Martin Drive, Baltimore, MD 21210, USA}

\author[0000-0001-9491-7327]{Simon P. Driver}
\affiliation{International Centre for Radio Astronomy Research (ICRAR) and the International Space Centre (ISC), The University of Western
Australia, M468, 35 Stirling Highway, Crawley, WA 6009, Australia}

\author{Connor Gelb}
\affiliation{School of Earth and Space Exploration, Arizona State University, Tempe, AZ 85287-1404, USA}

\author{Zak Goisman}
\affiliation{School of Earth and Space Exploration, Arizona State University, Tempe, AZ 85287-1404, USA}

\author[0000-0001-9440-8872]{Norman Grogin}
\affiliation{Space Telescope Science Institute, 3700 San Martin Drive, Baltimore, MD 21210, USA}

\author[0000-0003-4563-8983]{Daniel Henningsen}
\affiliation{School of Earth and Space Exploration, Arizona State University, Tempe, AZ 85287-1404, USA}

\author[0000-0002-4031-6400]{Isabela Huckabee}
\affiliation{School of Earth and Space Exploration, Arizona State University, Tempe, AZ 85287-1404, USA}

\author[0000-0003-0214-609X]{Scott J. Kenyon}
\affiliation{Smithsonian Astrophysical Observatory, 60 Garden Street, Cambridge, MA 02138, USA}

\author[0000-0002-6610-2048]{Anton M. Koekemoer}
\affiliation{Space Telescope Science Institute, 3700 San Martin Drive, Baltimore, MD 21210, USA}

\author[0000-0003-0238-8806]{Darby Kramer}
\affiliation{School of Earth and Space Exploration, Arizona State University, Tempe, AZ 85287-1404, USA}

\author[0000-0001-6529-8416]{John Mackenty}
\affiliation{Space Telescope Science Institute, 3700 San Martin Drive, Baltimore, MD 21210, USA}

\author[0000-0003-0429-3579]{Aaron Robotham}
\affiliation{International Centre for Radio Astronomy Research (ICRAR) and the International Space Centre (ISC), The University of Western
Australia, M468, 35 Stirling Highway, Crawley, WA 6009, Australia}

\author{Steven Sherman}
\affiliation{School of Earth and Space Exploration, Arizona State University, Tempe, AZ 85287-1404, USA}

\received{October 13, 2022}
\revised{March 24, 2023}
\accepted{April 12, 2023}

\submitjournal{AJ}

\begin{abstract}

The diffuse, unresolved sky provides most of the photons that the Hubble Space Telescope (HST) receives, yet remains poorly understood. HST Archival Legacy program SKYSURF aims to measure the 0.2--1.6 \micron{} sky surface brightness (sky-SB) from over 140,000 HST images. We describe a sky-SB measurement algorithm designed for SKYSURF that is able to recover the input sky-SB from simulated images to within 1\% uncertainty. We present our sky-SB measurements estimated using this algorithm on the entire SKYSURF database. Comparing our sky-SB spectral energy distribution (SED) to measurements from the literature shows general agreements. Our SKYSURF SED also reveals a possible dependence on Sun angle, indicating either non-isotropic scattering of solar photons off interplanetary dust or an additional component to \newnew{Zodiacal Light}. \newnew{Finally, we update Diffuse Light limits in the near-IR based on the methods from \citet{carleton_2022}, with values of 0.009 \MJysr{} (22 \nWmsr{}) at 1.25 \micron{}, 0.015 \MJysr{} (32 \nWmsr{}) at 1.4 \micron{}, and 0.013 (25 \nWmsr{}) \MJysr{} at 1.6 \micron{}.} These estimates provide the most stringent all-sky constraints to date in this wavelength range. SKYSURF sky-SB measurements are made public on the official SKYSURF website and will be used to constrain Diffuse Light in future papers.

\end{abstract}



\section{Introduction} \label{sec:intro}

The diffuse sky is an extended source of light present in all astronomical images. It is responsible for the majority of all photons the Hubble Space Telescope (HST) receives \citep[e.g.,][]{windhorst_2022}, and yet remains one of the most challenging sources of light to study. The light from astrophysical sources that produce the diffuse sky are easily contaminated by several polluting sources of light. These components can be largely described by 1) light scattered and re-emitted from the Earth's surface (Earthshine), 2) the extended point spread function (PSF) of bright sources both in and out of the field of view and 3) instrumental effects \citep{borlaff_2019}. In this work, we present a novel method to isolate the diffuse sky surface brightness (sky-SB) from $0.2-1.6$ \micron{} in over 140,000 HST images as part of the SKYSURF archival legacy program. Using this method, we present the first comprehensive study of the sky-SB as observed by HST using the entire Hubble Legacy Archive. 

The measured sky-SB in HST images is a combination of both astrophysical sources and polluting stray light. Zodiacal Light (ZL) is the brightest of the astrophysical sources and is produced by the scattering and re-emitting of Sunlight from interplanetary dust particles concentrated in the inner Solar System \citep[][]{hulst_1947, leinert98, kelsall98, sano_2020, korngut_2022}. ZL is so bright that it can account for more than 95\% of the sky-SB at $1.25$ \micron{}. Another component of the diffuse sky is the Diffuse Galactic Light (DGL), which is light scattered by dust and gas in the interstellar medium, as well as unresolved faint starlight. Finally, the faintest astrophysical contribution to the diffuse sky is the Extragalactic Background Light (EBL) consisting of \textit{all} far-UV to far-IR extragalactic photons, including light from stars, AGN, and dust attenuation/ re-radiation \cite[e.g.,][]{andrews_2017, hill_2018, driver_2021}.

\newnew{One key goal of SKYSURF is to provide a comprehensive archive of sky-SB measurements to improve existing ZL models. The ZL contains no resolvable structural component or strong spectral features from which to isolate it from other components of the sky-SB in HST images. In addition, the Earth orbits inside the IPD, making ZL especially difficult to constrain from Low Earth Orbit as there is no direction in which a space telescope can point to avoid it. Therefore, SKYSURF and many other programs rely on models of the Interplanetary Dust Cloud (IPD) from historical observations of the sky-SB in order to isolate it from the DGL and EBL. However, ongoing discrepancies in ZL models make studies of the other components of the sky-SB difficult \citep[e.g.,][]{korngut_2022}.}

\newnew{The \cite{kelsall98} ZL model is widely used, as it was the first ZL model to utilize NASA's Cosmic Background Explorer (COBE) Diffuse Infrared Background Experiment (DIRBE). It characterizes the annual modulation of ZL emission to produce a three-dimensional model, covering a broad spectral range from 1.25 \micron{} to 240 \micron{}.} COBE/DIRBE ZL emission maps have excellent relative accuracy of 1\% to 2\% \citep[][]{leinert98}, but are limited to Sun angles of 94 deg $\pm$ 30 deg. \new{In this work we define Sun angle to be the angle between the observation and the Sun, as shown in Figure 2 of \citeauthor{caddy_2022} \citeyear{caddy_2022}. This range of Sun angles would allow the detection of any nearby spherically symmetric component of the IPD \citep[][]{hauser_1998}, where Sun angles furthest from 90\degree{} will show the greatest variation in brightness as a function of ecliptic latitude. This limited range of Sun angles would potentially miss a more distant spherical component \citep[e.g.,][]{sano_2020}, where the relative changes in brightness would be smaller. This more distant component would appear to be isotropic.} In contrast, \citet{wright98} used COBE/DIRBE data with the condition that the faintest 25 \micron{} sky-SB at high ecliptic latitudes is only ZL. This condition means that the \citet{wright98} model includes flux from \text{any} isotropic component that the \citet{kelsall98} model might not account for, but also runs the risk of attributing some EBL and DGL to ZL. In addition to these infrared ZL models, \cite{leinert98} introduced a parametric ZL model that incorporates optical wavelengths ($0.2-0.9$ \micron) with limited data. This model assumes ZL to follow a reddened solar spectrum, which was later modified by \cite{aldering_2001} based on observations at the North Ecliptic Pole.

\newnew{In this work we present sky-SB measurements covering a more comprehensive range of Sun angles to build upon existing near infrared ZL models. We also present the first archive of sky-SB measurements with comprehensive optical sky coverage, which will help to better constrain ZL emission at these wavelengths. Quantifying the component of DGL in SKYSURF images is outside the scope of this work and will be addressed as part of future work.}

Another key goal of SKYSURF is to derive robust upper limits on the EBL. A secure measurement of the EBL is vital to deriving constraints on galaxy formation and evolution \citep[e.g.,][]{dominguez_2011, somerville_2012}, as it as probes star formation, AGN activity, and dust properties over cosmic time \citep[e.g.,][]{andrews_2017}. However, due to the challenges involved in measuring the sky-SB that we have discussed, it remains the least understood component of the sky-SB \citep[e.g.,][]{hill_2018, driver_2021}. Direct EBL measurements \citep[e.g.,][]{hauser_1998, dwek_1999,cambresy_2001, matsumoto_2005, bernstein_2007, dole_2006, matsuura_2017, lauer_2021} require robust subtraction of foregrounds such as ZL and DGL, and absolute calibration of the instrument. Some experiments use unique methods to better account for ZL emission. The CIBER experiments \citep[][]{matsuura_2017, korngut_2022} use the calcium triplet absorption feature in the ZL spectrum to better estimate the intensity of the ZL. The Pioneer \citep{matsumoto_2018} and New Horizons missions \citep{lauer_2021, lauer_2022, symons_2022} leave the inner Solar System entirely to heliocentric distances where ZL emission is systematically reduced, providing improvements ($>2\sigma$ significance) in direct measurements of EBL.

In contrast, an estimate of EBL can also be obtained by integrating the total flux from galaxy counts in deep surveys \citep[e.g.,][]{driver16, koushan_2021, windhorst_2023}. However, an interesting disparity is revealed when direct measurements of EBL are compared to these models derived from deep galaxy counts. This comparison yields 3--5$\times$ more EBL at optical wavelengths than we would expect based on galaxy counts alone \citep[see][]{driver16}. We refer to this unaccounted-for signal as Diffuse Light (DL), as its origin is unknown.

There are many potential sources for DL, including missing galaxies \citep{conselice_2016, lauer_2021}, the extended outskirts of galaxies \citep[e.g.,][]{li_2022, gilhuly_2022}, intrahalo light \citep{bernstein_1995, rudick_2011, mihos_2019}, reionization \citep[][]{santos_2002, cooray_2004, kashlinsky_2004, madau_2005, matsumoto_2011}, underestimated ZL or DGL models \citep{kawara_2017, korngut_2021}, telescope glow \citep{carleton_2022}, and Earthshine \citep{caddy_2022}, as well as more exotic sources such as dark matter particles \citep{maurer_2012, gong_2016} or black holes \citep{yue_2013}. See \citet{windhorst_2022} for a detailed summary.

\begin{figure*}[t]
    \centering
    \includegraphics[scale = 0.26]{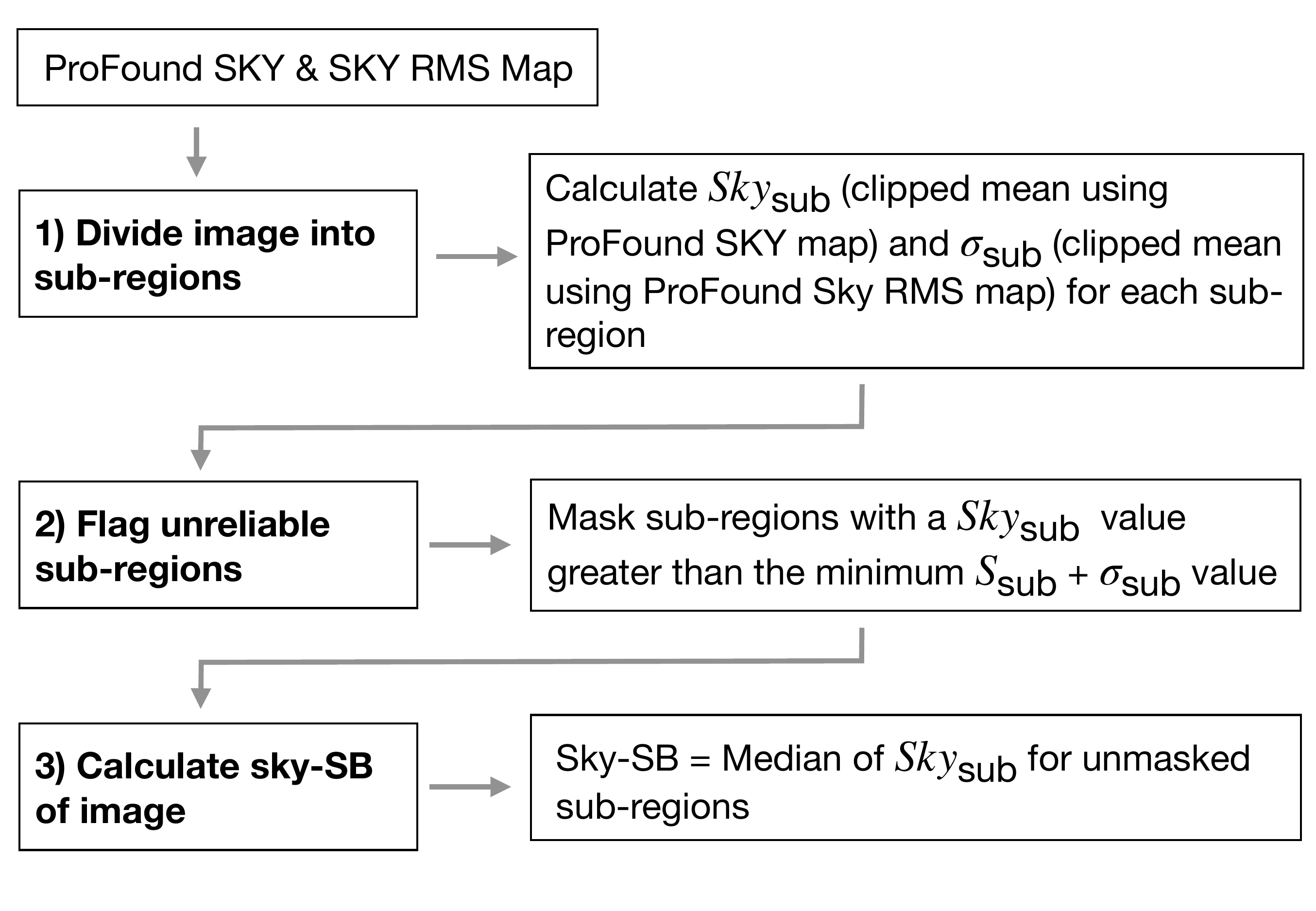}
    \caption{\new{Flowchart of the ProFound Median (Pro-med) pipeline to estimate the sky-SB from an HST image.}}
    \label{fig:methods_flowchart}
\end{figure*}

\begin{figure}[t]
    \centering
    \includegraphics[scale = 0.23]{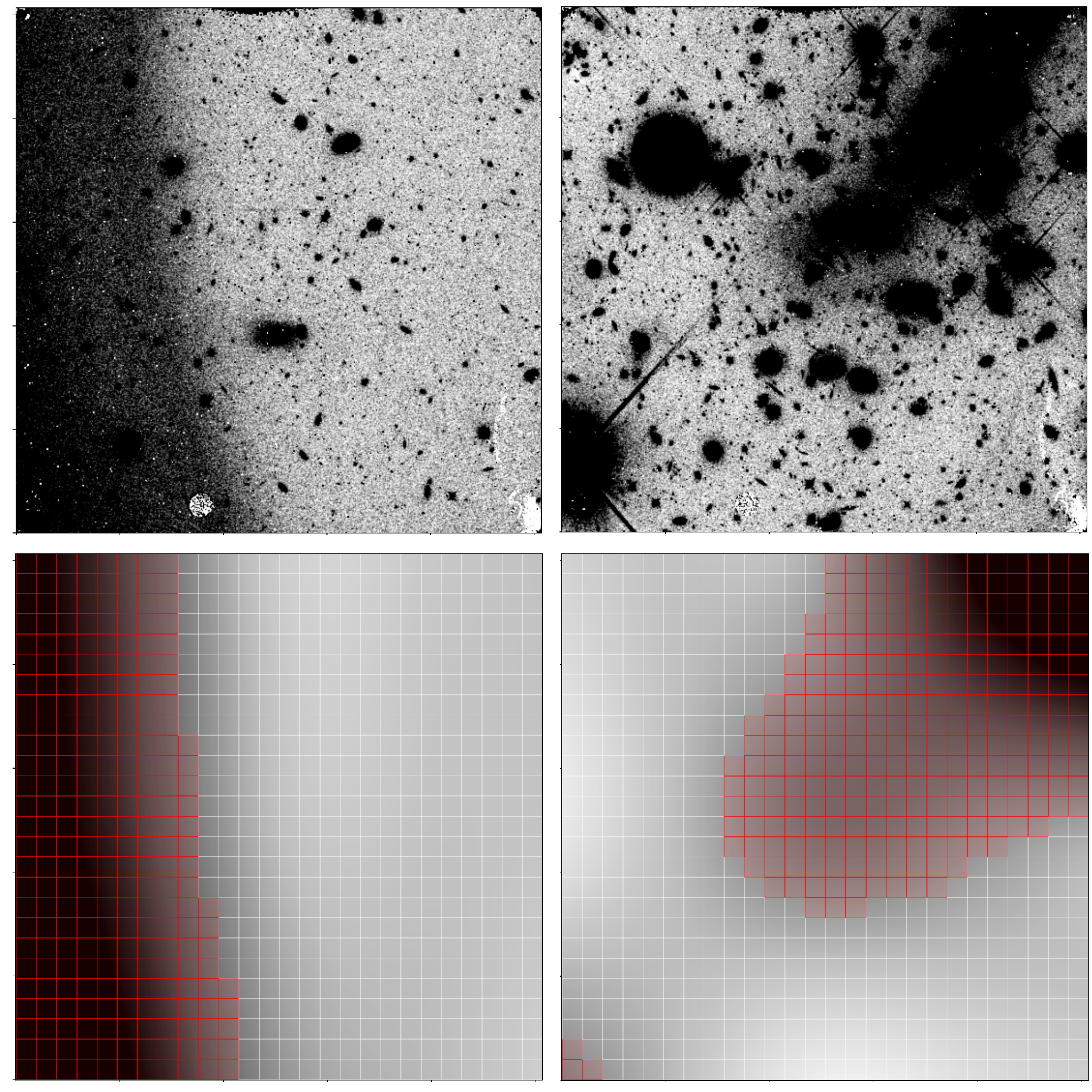}
    \caption{\new{SKYSURF sky-SB estimation algorithms are able to mask areas of an image containing light from discrete objects or bright sky gradients. The top row shows two example images in our SKYSURF database. The bottom row shows corresponding ProFound SKY maps, with the borders of Pro-med sub-regions overlaid. The bottom left ProFound map corresponds to the top left image, and similarly for the right-most images. The sub-regions used for the Pro-med algorithm are overlaid with white or red borders. Red sub-regions are identified by our algorithms as likely containing light contamination, and are masked during sky-SB estimation. We show these images to demonstrate our ability to reject areas of an image, but emphasize that images like this are rejected for sky-SB analysis because they are too crowded, as identified by the amount of red sub-regions.}} \label{fig:inspection_plots}
\end{figure}


The first step to better understanding the unaccounted for DL is reliable and robust sky-SB measurements. HST's capability as an ultra-sensitive, absolute photometer provides us with the necessary sensitivity to study the sky-SB. As a a result, our sky-SB measurements will enable the creation of improved, robust ZL models, and aid in constraining the amount of EBL and DL in the universe. 

\newnew{\citet[][hereafter SKYSURF-1]{windhorst_2022} summarizes Project SKYSURF, and \citet[][hereafter SKYSURF-2]{carleton_2022} provide in a pilot study upper limits to DL at three near-infrared wavelengths. This paper presents our methods and results for attaining robust panchromatic sky-SB measurements for all filters and all images in our SKYSURF database.} In Section \ref{sec:database}, we describe our SKYSURF database in detail. Section \ref{sec:methods} then explains our methods for estimating the sky-SB from any HST image. In Section \ref{sec:sky_vs_wave_results}, we show our main sky-SB results across our entire database, including a SKYSURF spectral energy distribution (SED) of the sky-SB. We present updated SKYSURF DL limits in Section \ref{sec:diffuse_light_limits}. Finally, Section \ref{sec:data} describes the SKYSURF data products that are released to the public.

\section{SKYSURF Database} \label{sec:database}

The SKYSURF database includes more than $140,000$ HST observations that are selected and summarized in Tables 1--3 of \citetalias{windhorst_2022}. In this paper, we focus on data taken with HST's newest cameras: the Wide Field Camera 3 (WFC3) and the Advanced Camera for Surveys (ACS). We utilize the UV-Optical (UVIS) and Infrared (IR) channels of WFC3, as well as the Wide Field Channel (WFC) of ACS. By utilizing different cameras with similar bandpasses we are able to identify how different instrument calibration processes might affect the sky-SB.

With such a large database, we expect a varied range of target types (e.g., gas clouds, star clusters, sparse fields, resolved galaxies, etc) and exposure types (e.g., short and long imaging exposures, grism finder images or DASH observations). In the scope of the SKYSURF program, certain exposures are not useful for the purpose of probing the sky-SB SED or are likely to contain high systematic uncertainties. These include very short exposures, exposures taken through narrowband filters, quad filters, polarizers, and dispersive elements. \new{Exposure times less than 200 seconds, where the read noise is relatively high, were not used. Experimentation with our sky measurement algorithms on simulated images showed that we were not able to get reliable sky measurements for these exposures.} In addition, the sky-SB is so low in UV images with short exposure times that errors associated with post-flash \citep[][]{biretta_2013} become a significant issue. The sky-SB measurement algorithms presented in Section \ref{sec:methods} are optimized for intermediate to long exposures through HST's sensitive wider band filters. We also do not utilize images taken in a subarrayed mode, where only a specific portion of the detector is used. Finally, ACS includes linear ramp filters (full WFC coverage at continuously varying narrow bandwidth) which are not used for SKYSURF.

The standard WFC3 and ACS pipelines create two main types of bias-subtracted, dark-frame subtracted, flat-fielded images: \flt{} and \flc{} files, where the latter includes Charge Transfer Efficiency (CTE) corrections. Since CTE trails do not affect non-destructively read near-IR detectors, we use the \flt{} files for WFC3/IR. We measure sky-SB levels on 143,231 WFC3 and ACS images. This includes 41,431 WFC3/IR \flt{} images, 26,542 WFC3/UVIS 
\flc{} images, and 75,258 ACS/WFC \flc{} images. Within this sample, there are 4,767 unique proposals and 22,883 unique target names. We report sky-SB measurements through 6 WFC3/IR, 14 WFC3/UVIS, and 8 ACS/WFC filters.

\section{Sky-SB Measurement Methods} \label{sec:methods}

In order to produce robust sky-SB measurements with HST, \newnew{we make use of sky-SB estimation algorithms designed for SKYSURF}, careful consideration of uncertainties, and the removal of stray light dominated images from our sample.

\subsection{Sky-SB Algorithms} \label{sec:sky_algorithms}

As shown in \citetalias{windhorst_2022} and Appendix \ref{sec:appendix_methods}, we tested various algorithms using simulated images. We simulate WFC3/IR F125W images with realistic galaxy and star counts. We explore a range of exposure times from 50 to 1302 seconds and sky-SB levels ranging from 0.22 to 3.14 electrons per second. \new{In order to simulate the effects of out of field stray light and ZL gradients in our test images, we apply sky gradients to our simulated images ranging from 0\% to 20\% across the field of view. ZL gradients will be much smaller than 20\% in a typical WFC3 or ACS field of view, but observations taken close to the Earth's limb may approach 20\%. Testing images that contain these steep gradients allows us to ensure our algorithms can consistently isolate the true sky-SB even under the most challenging conditions.} Appendix \ref{sec:simulations} describes the creation of these simulated images in detail.

We initially choose the two algorithms which are able to consistently retrieve the known input sky-SB from the simulated images: the ProFound Median (Pro-med) method and the ProFound-5th (Pro-5th) method. \new{Although these algorithms were developed independently, the Pro-med and Pro-5th algorithms are very similar, aside from the fact that the former uses a median and the latter uses a 5th-percentile.} Nevertheless, both are found to be able to retrieve the input sky-SB to within 1\% (see Figure \ref{fig:methods}). \new{For this paper, we focus on the Pro-med algorithm, which is found to be most accurate in general. Nevertheless, Appendix \ref{app:5th_method} explains the Pro-5th method in detail, as well as details sky-SB measurements using this algorithm.}

A flowchart showing our pipeline is given in Figure \ref{fig:methods_flowchart}. We utilize sky-SB maps created with ProFound \citep{robotham_2018}, a source finding and image analysis package that is able to interpolate behind objects to create a map of the sky-SB. To do this, ProFound utilizes a discrete boxcar filter on a grid. The resulting coarse grid is bicubic interpolated to construct the sky-SB map. Any objects detected in the image are masked. We refer to these ProFound sky-SB maps specifically as ProFound SKY maps herein. An example of ProFound SKY maps are shown in the bottom row of Figure \ref{fig:inspection_plots}.

We create ProFound SKY maps for all images in the SKYSURF database. We run ProFound with a large \texttt{box} size of 1/3 of the image dimensions, which decreases the resolution of the SKY maps and helps smooth over any local effects from bright objects that can leave behind features. To best remove light from the outskirts of objects, we enlarge the object masks such that the sky-SB around the enlarged aperture is correct. In addition, we mask every pixel flagged in the image data quality extension, as well as its immediate neighboring pixels. ProFound was run using default parameters otherwise.

\begin{table*}
\centering
\begin{tabular}{|cc|cc|}
\hline
  Camera & Filter &  Reliable Images [\#] & Reliable Images [\%]\\
\hline
  ACS/WFC &  F435W &                              1664 &                               15 \\
  ACS/WFC &  F475W &                               915 &                                8 \\
  ACS/WFC &  F555W &                               196 &                                4 \\
  ACS/WFC &  F606W &                              4817 &                               15 \\
  ACS/WFC &  F625W &                               373 &                               12 \\
  ACS/WFC &  F775W &                              1888 &                               10 \\
  ACS/WFC &  F814W &                              8918 &                               16 \\
  ACS/WFC & F850LP &                              5013 &                               29 \\
\hline
WFC3/UVIS &  F225W &                               346 &                               15 \\
WFC3/UVIS &  F275W &                               684 &                                8 \\
WFC3/UVIS &  F300X &                                52 &                               18 \\
WFC3/UVIS &  F336W &                               568 &                                7 \\
WFC3/UVIS &  F390W &                               347 &                               19 \\
WFC3/UVIS &  F438W &                               118 &                                5 \\
WFC3/UVIS &  F475X &                                42 &                                6 \\
WFC3/UVIS &  F475W &                               289 &                               15 \\
WFC3/UVIS &  F555W &                                62 &                                2 \\
WFC3/UVIS &  F606W &                              1138 &                               10 \\
WFC3/UVIS &  F625W &                                61 &                                7 \\
WFC3/UVIS &  F775W &                                43 &                                7 \\
WFC3/UVIS & F850LP &                                68 &                               17 \\
WFC3/UVIS &  F814W &                              1241 &                                9 \\
\hline
  WFC3/IR &  F098M &                               205 &                               18 \\
  WFC3/IR &  F105W &                              1102 &                               23 \\
  WFC3/IR &  F110W &                               497 &                                7 \\
  WFC3/IR &  F125W &                              1109 &                               20 \\
  WFC3/IR &  F140W &                               768 &                               16 \\
  WFC3/IR &  F160W &                              2460 &                               12 \\
\hline
\end{tabular}
\caption{Fraction of images with reliable sky-SB measurements for every SKYSURF filter. These images are chosen using the methods of Section \ref{sec:choosing_reliable_skys}. Reliable images have no more than 30\% of sub-regions flagged, contain expected noise levels based on Gaussian and Poisson noise, are not manually flagged, and are not significantly affected by persistence. \new{In addition, we remove images that are within 20\degree{} of the galactic plane, have sun altitudes greater than 0\degree, are too close to a large nearby galaxy, have Sun angles less than 80\degree, or have moon angles less than 50\degree.} We list the number of reliable sky-SB measurements and the percent of total images in the SKYSURF database that are reliable.} \label{tab:fraction_reliable}
\end{table*}

\begin{table*}[t]
    \centering
    \begin{tabular}{|c|c|c|c|}
    \hline
    Uncertainty       & WFC3/UVIS                  & WFC3/IR                       & ACS/WFC  \\
    \hline
    Sky Algorithm     & 0.4\%                      & 0.4\%                         & 0.4\%      \\
    Flat-field        & 1\%                        & 2\%                           & 2.2\%             \\
    Zeropoint         & 0.2\%                      & 1.5\%                         & 1\%               \\
    Non-linearity     & N/A                        & 0.5\%                         & N/A               \\
    \hline
    Bias              & 0.2 $e^-$ (1.4\%)          & 0.005 $e^-$/s (0.7\%)         & 0.6 $e^-$ (1.5\%)       \\
    Dark              & 0.3 $e^-$ (2.1\%)          & 0.005 $e^-$/s (0.7\%)         & 0.5 $e^-$ (1.2\%)        \\
    Thermal Dark      & N/A                        & 0.01 $e^-$/s  (1.3\%)         & N/A               \\
    Post-flash        & 0.16 $e^-$ (1.1\%)         & N/A                           & 0.37 $e^-$ (0.9\%)       \\
    \hline
    Total             & 3.0\%                      & 3.1\%                         & 3.2\%       \\
    \hline
    \end{tabular}
    \caption{SKYSURF sky-SB uncertainties. Multiplicative uncertainties (top rows) are listed as a percent of the sky-SB for WFC3/UVIS, WFC3/IR, and ACS/WFC. Additive uncertainties (bottom rows) list the error in units of electrons ($e^-$) or electrons per second ($e^-$/s), \new{with the average percent of the sky-SB shown in parenthesis}. Sky Algorithm refers to the ability of our algorithm to retrieve the true input sky from simulated images, and follows from \citetalias{windhorst_2022}. Flat-field refers to uncertainties in flat-field correction. Zeropoint refers to uncertainties in detector zeropoints. Non-linearity refers to the non-linearity of WFC3/IR. Bias and Dark refer to subtraction uncertainties in bias and dark frames. Thermal Dark refers to uncertainties in the Thermal Dark signal (thermal noise from the telescope assembly and instruments) described in \cite{carleton_2022}. Post-flash refers to uncertainties is post-flash subtraction. \new{The last row shows the total error as a percent of the sky-SB, which is a median uncertainty of all images in the corresponding camera.} \\
    \textbf{WFC3/UVIS:} Flat-field \citep{mack_2016}, Zeropoint \citep{calamida_2022}, Bias \citep{mckay_2017}, Dark \citep{bourque_2016}, Post-flash \citepalias{windhorst_2022} \\
    \textbf{WFC3/IR:} Flat-field \citep{mack_2021}, Zeropoint \citepalias{windhorst_2022}, Non-linearity \citepalias{wfc3_ihb}, Bias \citepalias{windhorst_2022}, Dark \citepalias{windhorst_2022}, Thermal Dark \citetalias{carleton_2022} \\
    \textbf{ACS/WFC:} Flat-field \citep{cohen_2020}, Zeropoint \citep{bohlin_2020}, Bias \citep{acs_dhb}, Dark \citep{anand_2022}, Post-flash \citepalias{windhorst_2022}
    }
    \label{tab:error}
\end{table*}

\begin{figure*}[t]
    \centering
    \includegraphics[scale = 0.44]{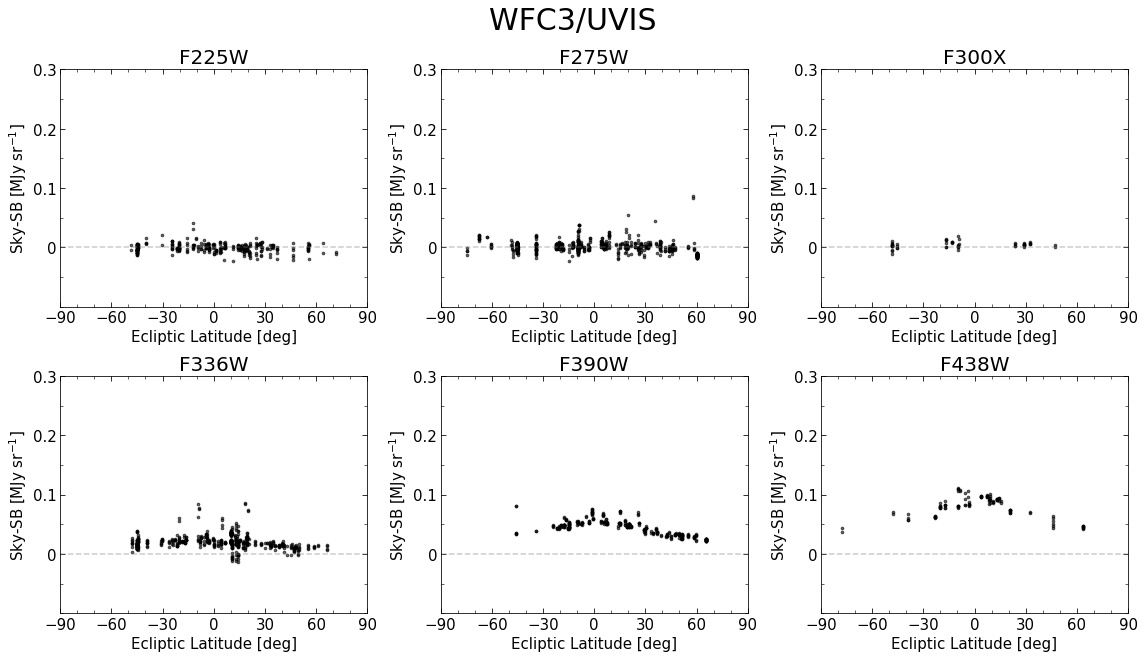}
    \caption{SKYSURF sky-SB measurements versus Ecliptic Latitude for WFC3/UVIS (UV filters), zoomed in to show the range between -0.1 to 0.3 \MJysr. }
    \label{fig:sky_vs_ecllat_wfc3uvis_1}
\end{figure*}

\begin{figure*}[t]
    \centering
    \includegraphics[scale = 0.44]{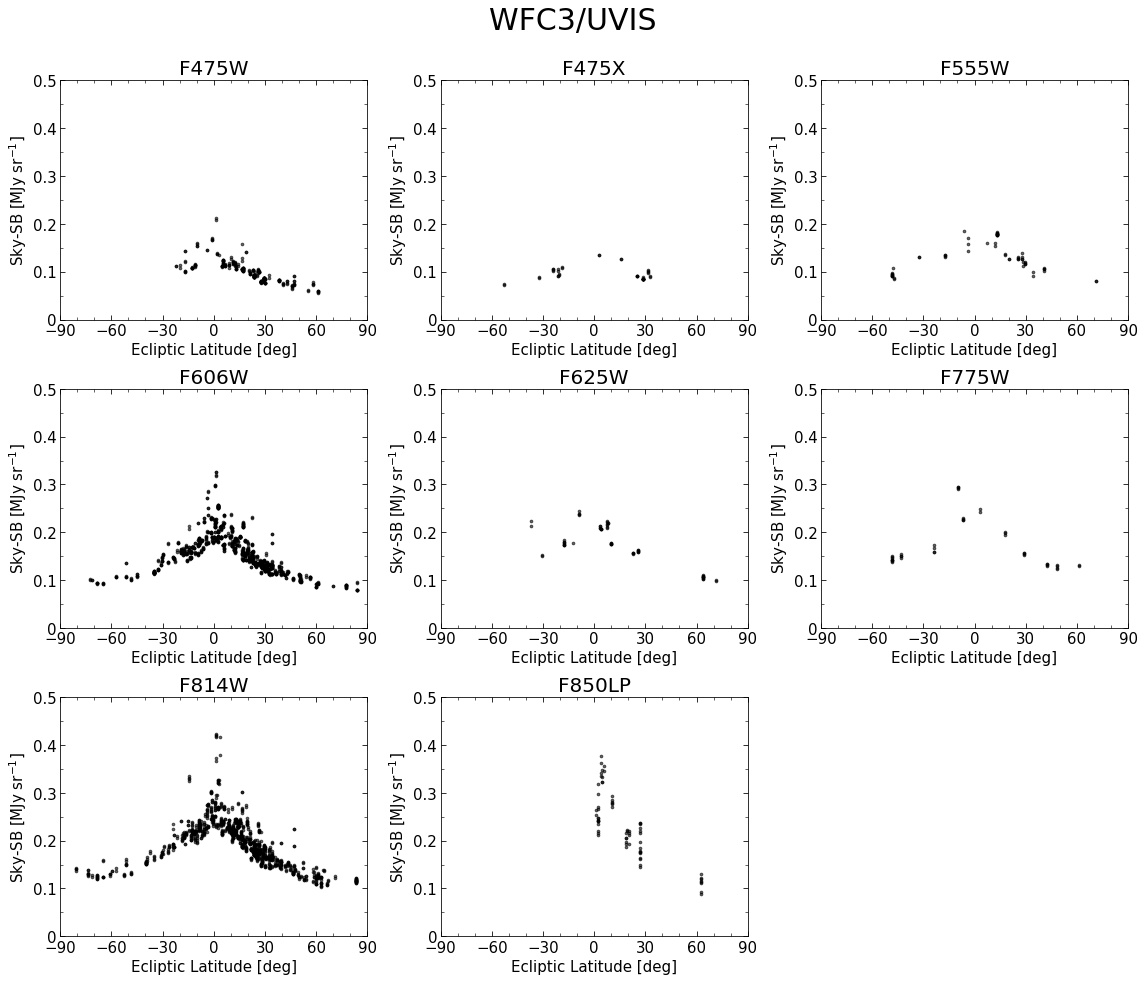}
    \caption{SKYSURF sky-SB measurements versus Ecliptic Latitude for WFC3/UVIS (Optical filters), zoomed in to show the range between 0 to 0.5 \MJysr. There is a larger scatter in the F850LP filter due to the high sky-SB rms of this filter.}
    \label{fig:sky_vs_ecllat_wfc3uvis_2}
\end{figure*}

\begin{figure*}[t]
    \centering
    \includegraphics[scale = 0.44]{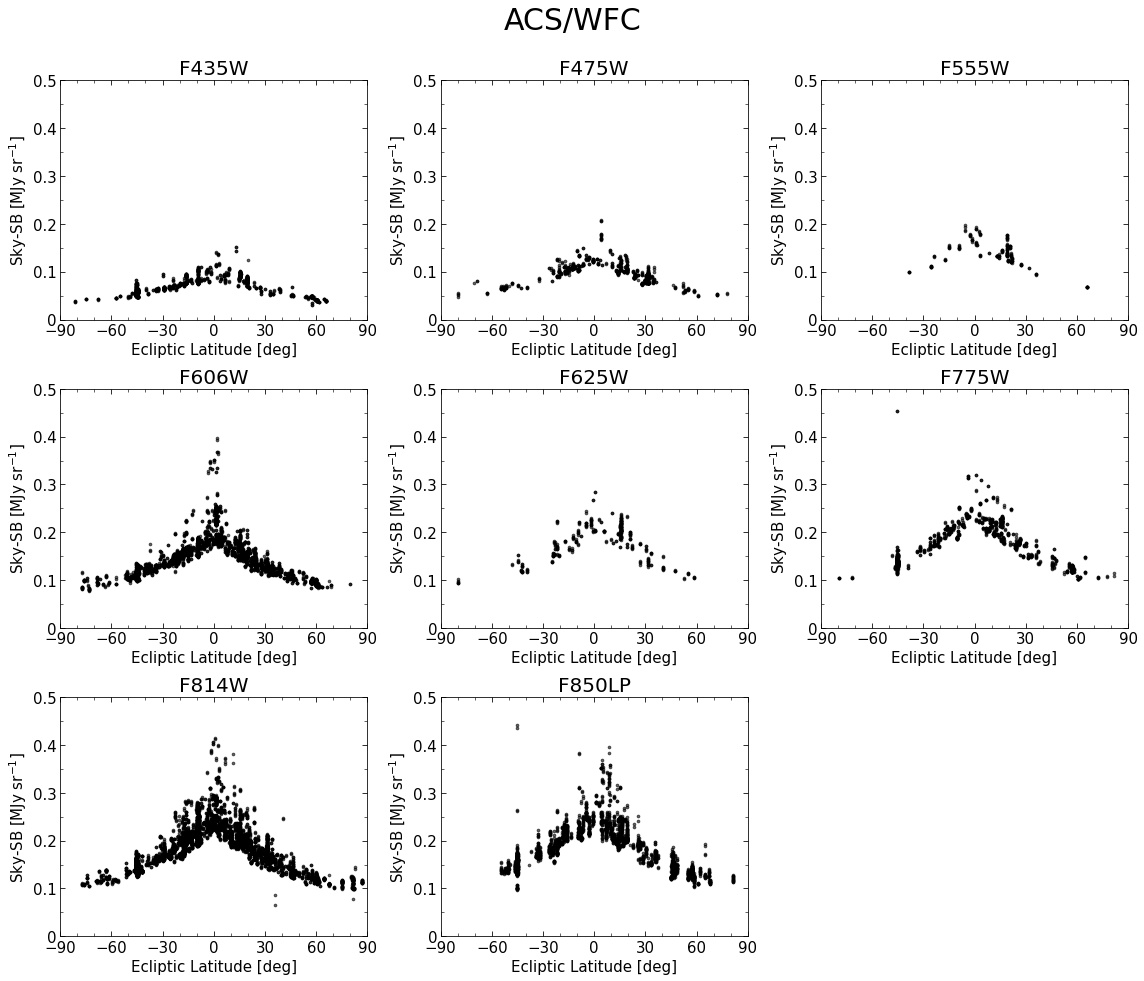}
    \caption{SKYSURF sky-SB measurements versus Ecliptic Latitude for ACS/WFC, zoomed in to show the range between 0 to 0.5 \MJysr.}
    \label{fig:sky_vs_ecllat_acswfc}
\end{figure*}

\begin{figure*}[t]
    \centering
    \includegraphics[scale = 0.44]{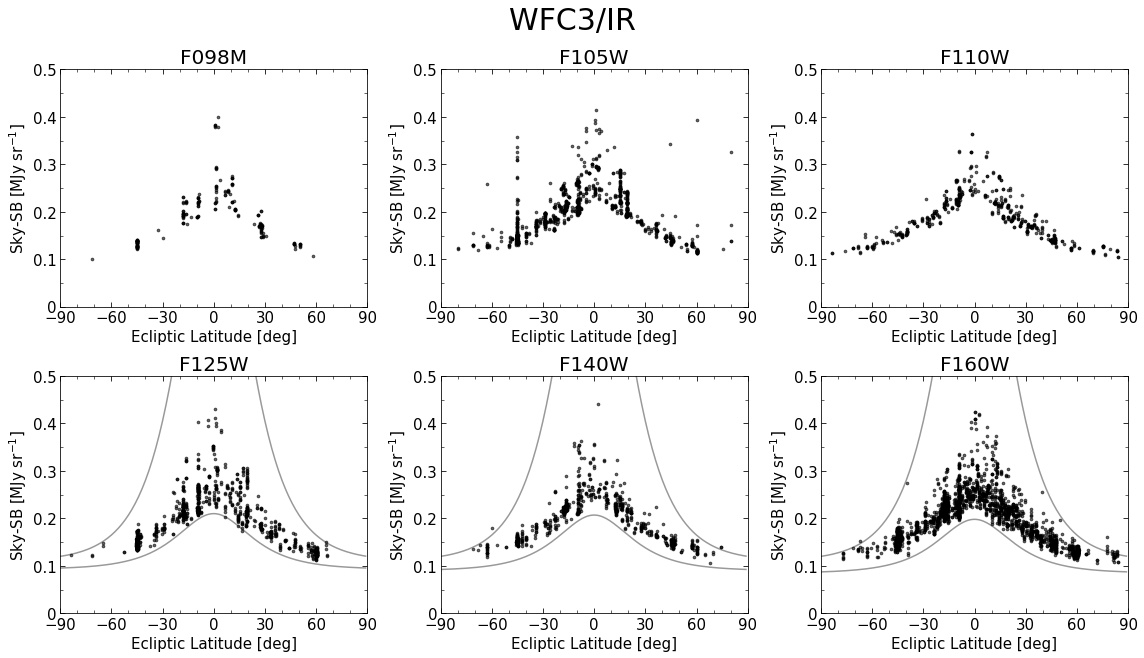}
    \caption{SKYSURF sky-SB measurements versus Ecliptic Latitude for WFC3/IR, zoomed in to show the range between 0 to 0.5 \MJysr. The large scatter in points in the F105W filter at $-45$\degree{} is due to observations of the Hubble Ultra Deep Field, where the 1.083 \micron{} emission line \citep{brammer_2014} present in this filter likely contaminated some of these measurements. \new{As an example of the sky-SB measurements we expect, we include the F125W, F140W, and F160W $sech$ curves from Figure 2 of \citetalias{carleton_2022}, which are derived to match \citet{kelsall98} Zodiacal model predictions.} Our measurements fall conservatively within these limits.}
    \label{fig:sky_vs_ecllat_wfc3ir}
\end{figure*}

\begin{figure*}[t]
    \centering
    \includegraphics[scale = 0.54]{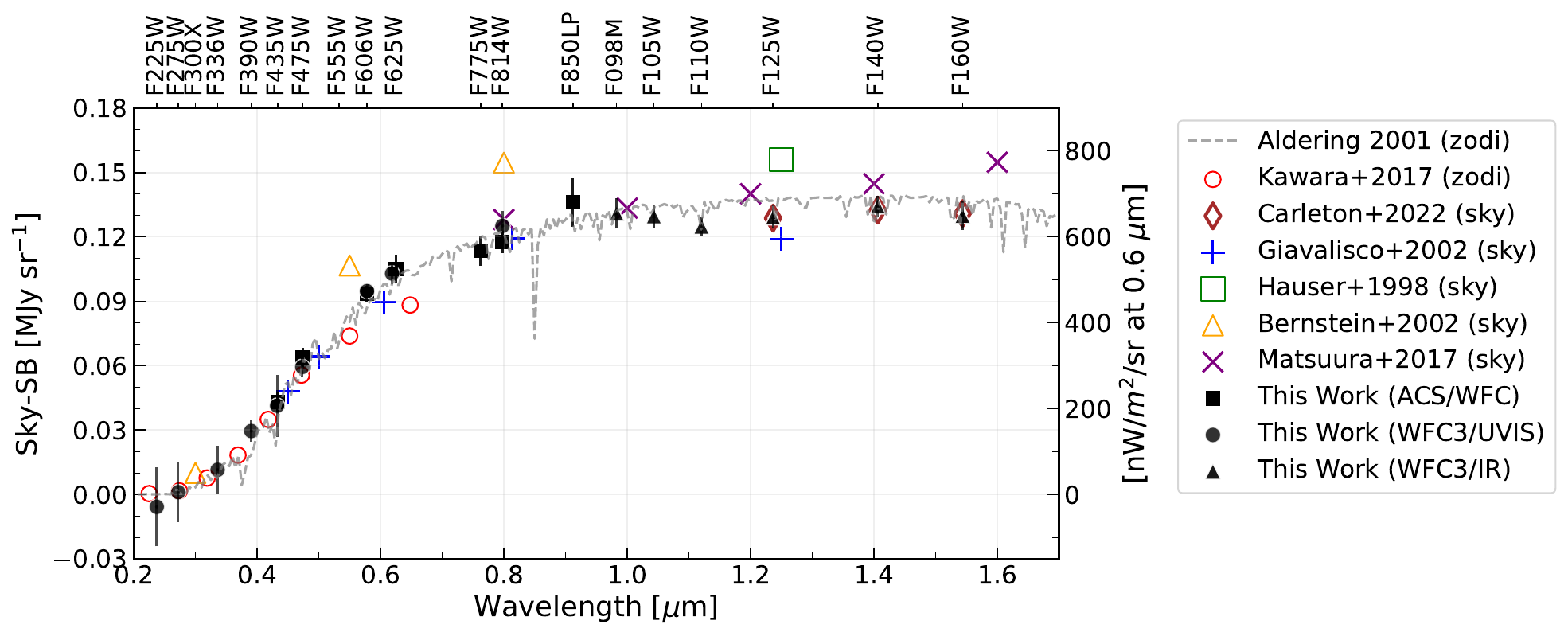}
    \caption{SKYSURF sky-SB SED for Ecliptic latitudes within 40\degree{} of the poles. Each black point is a ACS/WFC (square), WFC3/UVIS (circle) or WFC3/IR (triangle) median sky-SB measurement. The error bar is the standard error in unique positions in the sky and the sky-SB measurement error, added in quadrature (see Equation \ref{eq:errorbar}). Only filters with at least two unique positions are plotted. We compare our measurements to comparable measurements of the sky-SB \new{taken near the ecliptic poles with space telescopes or sounding rockets} \citep{giavalisco_2002, hauser_1998, bernstein_2002, matsuura_2017, carleton_2022}, a model of ZL emission \citep{aldering_2001}, and measurements of ZL emission \citep{kawara_2017}. \new{Overall, SKYSURF sky-SB measurements agree well with other models and predictions, although some offsets remain present in studies where the light from discrete objects is not removed \citep{hauser_1998, bernstein_2002, matsuura_2017}.}}
    \label{fig:sky_vs_wave_pole}
\end{figure*}

\begin{figure*}[t]
    \centering
    \includegraphics[scale = 1]{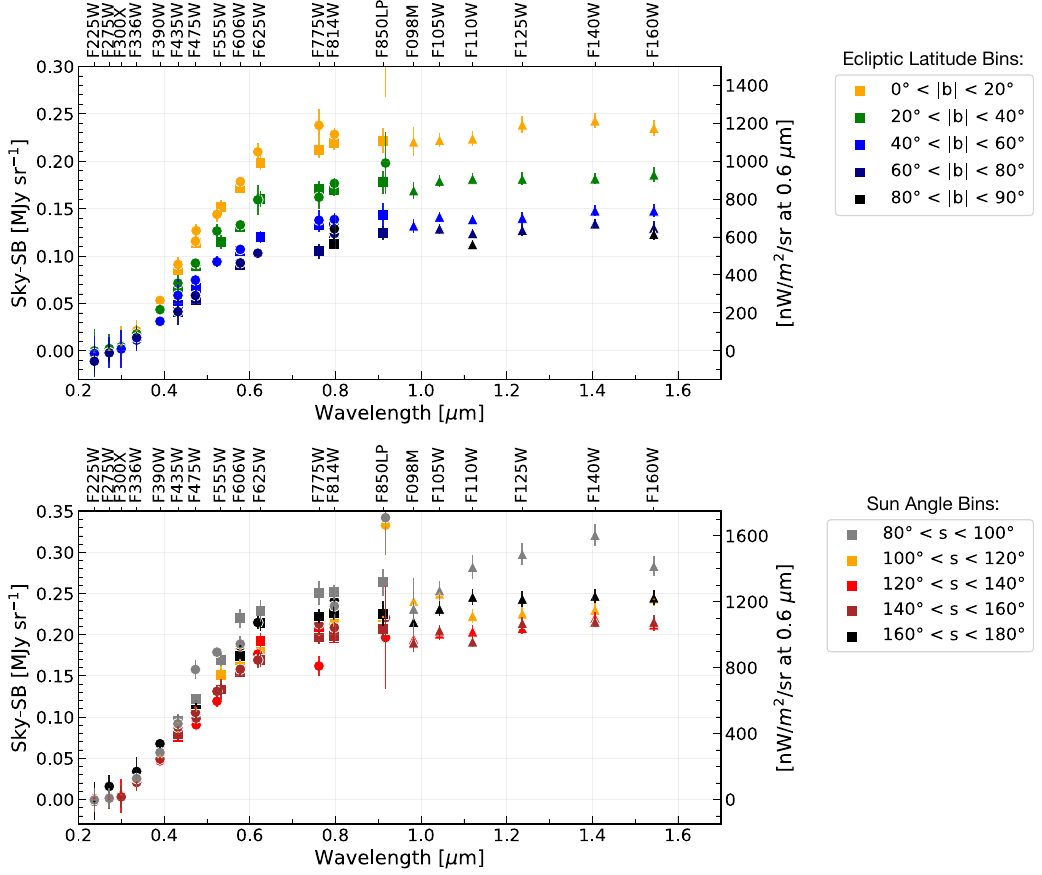}
    \caption{SKYSURF sky-SB SED for different Ecliptic Latitude ($b$, where each bin includes the absolute value of $b$) and Sun angle ($s$) bins. \new{Since Sun angle can be correlated with Ecliptic Latitude, the bottom plot has a fixed range in Ecliptic Latitude of $-30\degree < b < 30\degree$.} Each point is a ACS/WFC (squares), WFC3/UVIS (circles) or WFC3/IR (triangles) 3$\sigma$-clipped median sky-SB measurement. The error bars are scaled in the same way described in Figure \ref{fig:sky_vs_wave_pole}.}
    \label{fig:sky_sed_ecllat_and_sunang}
\end{figure*}

\begin{figure*}[t]
    \centering
    \includegraphics[scale = 0.6]{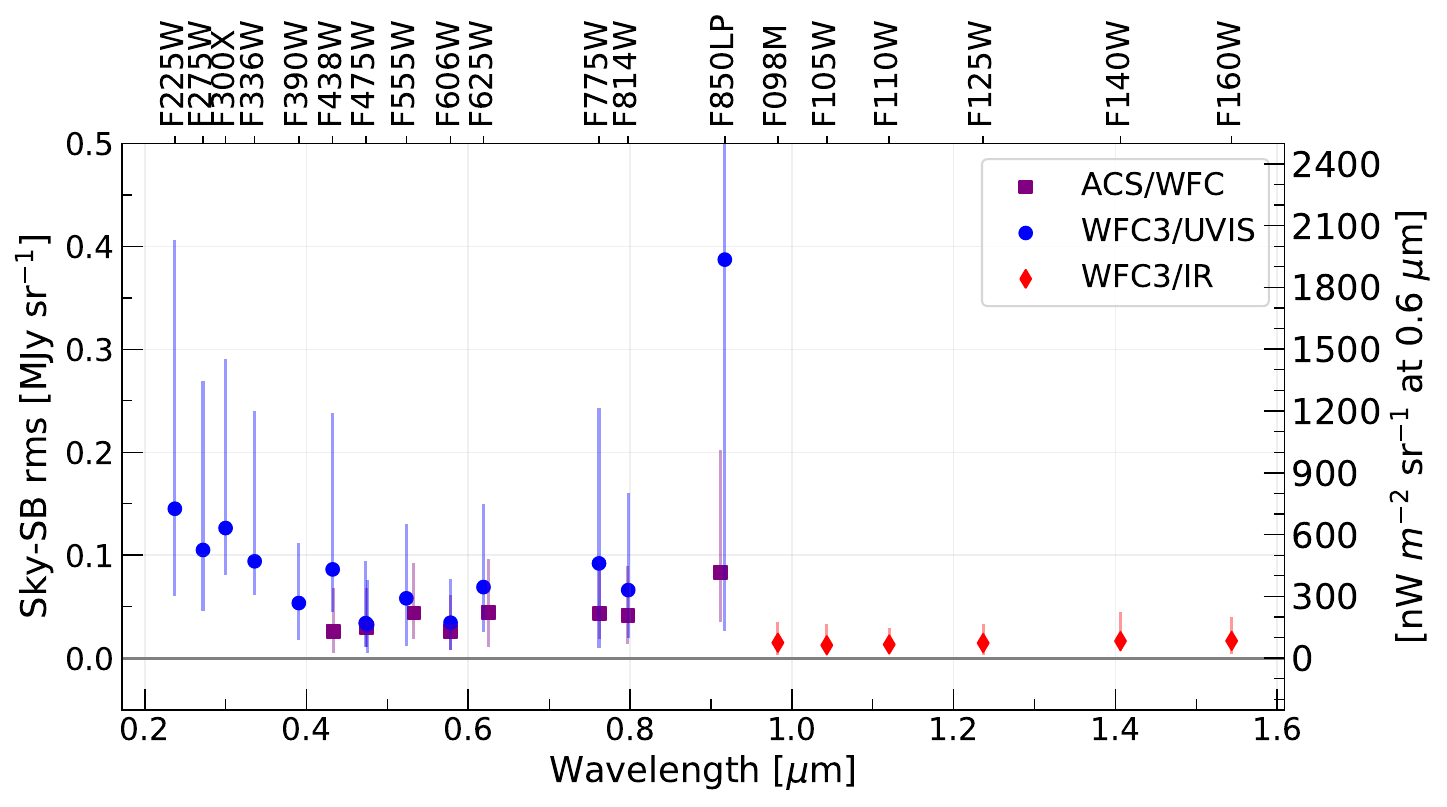}
    \caption{Median sky-SB rms for each filter. The error bars show the 16th- and 84th-percentiles of the rms distributions. \new{The sky-SB rms trends seen here are likely due to the different pixel sizes of each camera.}}
    \label{fig:rms_vs_wave}
\end{figure*}

The Pro-med method follows three steps: 1) divide each ProFound SKY map into sub-regions and calculate sky-SB levels for each sub-region, 2) flag sub-regions with unreliable sky measurements, and 3) take the median of unflagged sub-regions to determine the sky-SB of the image. All WFC3/IR images in our sample are $1014\times1014$ pixels, and we create $39\times39$ pixel sub-regions, for a total of $26\times26=676$ sub-regions per SKY map. For WFC3/UVIS, a full-frame detector image has dimensions of $2051\times4096$ pixels. For this case, we make each sub-region $64\times64$ pixels, where the few remaining pixels will be added to the sub-regions on the top-most row and right-most columns. For ACS/WFC, the detector has dimensions $2048\times 4096$, and we follow the same binning technique as for WFC3/UVIS. Both WFC3/UVIS and ACS/WFC images include two CCD detectors (stored as separate science extensions), and we perform sky-SB measurements on each detector independently.

Next, we calculate the sky level ($\skysub$) for each sub-region in native pixel units: electrons per second for WFC3/IR and electrons for WFC3/UVIS and ACS/WFC. We first mask all outlier pixels within a single sub-region that are not part of the background using boundary values of $-126.5$ and $895.5$, which were motivated by extremes in noisy WFPC2 data and are designed to eliminate the most obvious outliers. The mean ($\skysub^{'}$) and standard deviation ($\skysubrms^{'}$) of the remaining pixels are then recorded. We then mask pixels which have a value less than $\skysub^{'}-5\times$($\skysubrms^{'}$) or more than $\skysub^{'}+3\times$($\skysubrms^{'}$). \new{We are more stringent with the upper cutoff since, for normal images, the distribution of pixels leans towards the positive side.} New values of $\skysub^{'}$ and $\skysubrms^{'}$ are calculated for the remaining pixels. This process is repeated until there are no outlier pixels remaining. The median value of the last iteration is saved as the $\skysub$ ($\skysubrms$) for that sub-region.

In order to mask sub-regions, we compare each $\skysub$ in a single sub-region to all the other $\skysub$ values in a single image. If a $\skysub$ value for a sub-region is greater than the minimum $\skysub$+$\skysubrms$ of all the sub-regions, then we conclude that the sub-region likely contains light contamination from an object and this sub-region is masked. This step is critical to ensuring we are only utilizing sub-regions with the highest probability of being stray light free. Figure \ref{fig:inspection_plots} shows examples of how this algorithm masks sub-regions on \flt/\flc{} files.

The final sky-SB level of a chip, $\skyim$, is the median of the remaining $\skysub$ values. To calculate $\skysubrms$, we use the same method used to calculate $\skysub$ but instead perform it using the ProFound SKY RMS map. As shown in \citetalias{windhorst_2022} and Appendix \ref{sec:appendix_methods} here, this algorithm is demonstrated to recover the simulated sky-SB with an error of $<0.4$\% for expected sky gradients of $<10$\% on simulated F125W images.  

\newnew{As shown in Appendix \ref{sec:appendix_methods}, the Pro-med method is demonstrated to be the best estimator of the sky-SB. Therefore, the Pro-med method is chosen to derive robust sky-SB measurements in this paper, and thus produce the SKYSURF SED of the sky-SB presented in Section \ref{sec:sky_vs_wave_results}.}

\newnew{To further confirm that our sky-SB measurements are not contaminated by the light from discrete objects, we check whether our sky-SB measurements increase with the number of stars and galaxies in an image. If the sky-SB increased with the amount of light from discrete objects, it could indicate that our sky-SB measurements are accidentally picking up the faint outskirts of galaxies or the extended PSFs from stars. \citetalias{windhorst_2022} performs this test for the F125W filter only and finds no increase in the sky-SB as a function of the amount of object light. We tested this also for F140W and F160W, and find the same results. Therefore, a Diffuse Light limit estimated using the sky-SB algorithms in this work is unlikely to come from the faint outskirts of galaxies.}


\newnew{We subtract from WFC3/IR estimates the thermal dark signal as described in \citetalias{carleton_2022}. These values are listed in Table \ref{tab:thermal_dark}.}

\begin{table}[h]
    \centering
    \begin{tabular}{|c|cc|}
    \hline
    & \multicolumn{2}{c|}{Thermal Dark} \\
    Filter & [e s$^{-1}$] & [\MJysr] \\
    \hline
    F098M & 0.0044  & 0.0023 \\ 
    F105W & 0.0044  & 0.0013 \\
    F125W & 0.0040  & 0.0012 \\
    F140W & 0.0201  & 0.0050 \\
    F160W & 0.0772  & 0.0308 \\
    \hline
    \end{tabular}
    \caption{Thermal Dark values (Carleton et al., in prep) that are subtracted from sky-SB measurements in this work.}
    \label{tab:thermal_dark}
\end{table}

\subsection{Flagging Unreliable Images} \label{sec:choosing_reliable_skys}

In addition to the broad filtering of images described in Section \ref{sec:database}, after performing sky-SB measurements we filter out images where the sky-SB level is deemed unreliable. This includes measurements where the image is dominated by bright objects, images with very high sky-SB rms levels, or images where a guide star was lost. These measurements are still available for public use, but are not used in SKYSURF analysis.

Within the scope of this work, the sky-SB cannot be measured from images that are saturated with bright sources. These include exposures that contain star clusters, galaxies that take up a large portion of the field-of-view, or steep sky gradients. Since our algorithms record the number of sub-regions that are flagged as unreliable, a high number of unreliable sub-regions indicates that the sky-SB must also be unreliable. We reject sky-SB measurements where more than 30\% of sub-regions are flagged as unreliable. These regions are shown as red in Figure \ref{fig:inspection_plots}.

A high sky-SB rms ($\skyrmsim$) can also indicate an unreliable measurement. We predict the sky-SB rms of an image to be a combination of expected Gaussian and Poisson noise (shot noise):

\begin{equation}
    \sigma_{\text{predicted}} = \frac{\sqrt{ R^2 + S_{\text{chip}} \times t}}{t},
\end{equation}

\noindent where $\sigma_{\text{predicted}}$ is the predicted sky-SB rms in electrons per second, $R$ is the detector readnoise in electrons, $S_{\text{chip}}$ is the sky-SB in electrons per second, and $t$ is the exposure time in seconds. We reject images where $\skyrmsim > 2\sigma_{\text{predicted}}$.

In IR detectors an afterglow remains in pixels that were saturated in previous exposures. This phenomenon is called persistence and is known to affect WFC3/IR images. Therefore, we need to ensure it does not influence our sky-SB measurements. The standard pipeline for IR images does not eliminate flux from persistence. However, the WFC3 team has developed a software that estimates the amount of persistence per \flt{} file. This produces pixel maps (labeled as \verb|_persist.fits|) that estimate the amount of internal and external persistence for a given image \citep[see][]{wfc3_dhb}. We masked pixels in 30 random F160W images with values greater than 0.005 electrons per second in the \verb|_persist.fits| file and reran our sky-SB algorithm. For most images, we did not find significant differences in the sky-SB level when masking pixels that are affected by persistence. We found $\sim$0.25\% differences in sky-SB for images where more than 1\% of the pixels are affected by persistence, where an affected pixel is defined to be one with Flux$_\mathrm{persistence} > 0.01$ electrons per second. We therefore reject images where more than 1\% of the pixels are affected by persistence.

\new{To reduce the probability that images may be contaminated by sources like Earth's limb or the Milky Way, we follow the methods described in \citetalias{windhorst_2022}. We reject images within 20\degree of the galactic plane to reduce contamination due to Galactic cirrus, and densely populated star fields. To reduce the impact of Earthshine, images with Sun altitudes greater than zero are also removed. Here we define Sun altitude as the mean angle of the Sun above the Earth's horizon over the duration of an observation. Finally, we reject images with Sun angles less than 80\degree and with Moon angles less than 50\degree. Section 4.3 and Appendix A.2.2 of \citetalias{windhorst_2022} show that Sun angles less than 80\degree{} can be problematic. The minimum sky-SB for Sun angles $\lesssim$80\degree{} is greater than for Sun angles $\gtrsim$80\degree, in part due to increasing ZL intensity closer to the Sun, although the main cause is not known. For example, it is possible that Sun angles $\lesssim$80\degree{} are affected by scattered sunlight.}

We carefully reject images taken too close to nearby galaxies and globular clusters that might take up HST's entire field of view. We utilize galaxies selected as listed in the \newnew{Third Reference Catalog of Bright Galaxies \citep[RC3,][]{de_Vaucouleurs_1991}}. We locate the positions of all galaxies larger than 2 arcmin in diameter, and reject images that are within 1$R_{\text{galaxy}}$ to the center of the object. \newnew{$R_{\text{galaxy}}$ is adopted directly from the RC3, and is half the apparent major axis isophotal diameter of a galaxy measured at a surface brightness level of 25.0 B-mag per square arcsecond\footnote{\url{https://heasarc.gsfc.nasa.gov/W3Browse/all/rc3.html}}.}

Critically, we opted to manually inspect our SKYSURF images to understand how well our algorithm performs and try and identify images that are not automatically flagged using the process described above. We flag images that are smeared (due to loss of a guide star), or are clearly saturated with objects. \newnew{For our filter with the largest number of images, ACS/WFC F814W,} $<0.2$\% of the images were flagged, confirming the above methods to reject unreliable images. We reject remaining images flagged during the manual inspection.

In addition, during the manual inspection, we flag images with visible satellite trails and artifacts like optical ghosts. These images are not rejected. However, to ensure image artifacts are not accidentally recorded as objects, the list of satellite trails and optical artifacts are recorded for consideration for star and galaxy counts.

Table \ref{tab:fraction_reliable} shows the percent of images in each SKYSURF filter that are not flagged by the methods described here and are thus deemed reliable. The number of reliable images per filter is typically between 5--20\%, however some filters exhibit lower reliability due to being used frequently in problematic observations (e.g., F555W is often used for stellar populations studies, and thus includes many observations of star clusters). We emphasize that many of the images rejected here are still beneficial for galaxy and star counts and reducing cosmic variance. As a result, while they are not used to derive sky-SB limits in this work they are not excluded from the SKYSURF database. 

\subsection{Sources of Uncertainty} \label{sec:error}

There are many sources of uncertainty during data reduction and calibration that are inconsequential for studies of discrete objects, but remain important when measuring the sky-SB. \citetalias{windhorst_2022} summarizes the main SKYSURF sources of error in detail. We summarize updated sources of uncertainty in Table \ref{tab:error}.

\new{We follow the methods of \citetalias{windhorst_2022} to add our uncertainties in quadrature. We present our results assuming all systematics have been accounted for through various WFC3 and ACS Instrument Science Reports \citep[e.g.,][]{wfc3_dhb, acs_dhb}. Therefore, SKYSURF sky-SB uncertainties represent \textit{random} uncertainties in the ability to determine a bias frame, dark frame, flat-field, etc. In addition, in \citetalias{windhorst_2022} explains that the two dominant sources of error, which are zeropoint and flat-field uncertainties, are independent, and thus can be added in quadrature.}

Uncertainties for each camera are presented separately. We are careful to distinguish between multiplicative errors, which depend on the level of the sky-SB, and additive errors, which are the same regardless of image brightness. Further explanation of these uncertainties, as well as an exploration of the effect of crosstalk, amplifier differences, and flat-field residuals on sky-SB measurements, are presented in Appendix \ref{sec:error_appendix}.

\section{Sky-SB Measurement Results} \label{sec:sky_vs_wave_results}

We perform sky-SB measurements using the Pro-med algorithm and show our results here. The process for converting our images to units of flux density is described in Appendix \ref{app:converting_to_flux_units}.

\subsection{Sky-SB versus Ecliptic Latitude}

Figures \ref{fig:sky_vs_ecllat_wfc3uvis_1}-\ref{fig:sky_vs_ecllat_wfc3ir} show our results as a function of Ecliptic Latitude. \new{Results for some duplicated filters are presented independently for each camera, as the filters will have a different bandpass when paired with each camera. In addition, given their subtle zeropoint differences, this provides an independent check on the sky-SB values with different instruments.}

\new{The bottom row of Figure \ref{fig:sky_vs_ecllat_wfc3ir} includes the hyperbolic secant curves derived in \citetalias{carleton_2022} using the \citet{kelsall98} Zodiacal model. They represent the range in sky-SB values expected as a function of ecliptic latitude. \citetalias{carleton_2022} did not apply the same aggressive filtering of images described in this work. Therefore, our measurements don't span as large of a range as the grey curves might suggest. Overall, these limits are largely driven by Sun angle, where lower Sun angles result in brighter sky-SB levels at the same ecliptic latitude. Therefore, the range in expected sky-SB levels is broader closer to the ecliptic plane due to the broader range of Sun angle's available here. Our measurements fall conservatively within these limits, showcasing our ability to successfully reject images from our database.}

Some notable deviations from the expected sky-SB are the spike in sky-SB values at an ecliptic latitude of $-45$\degree{} in F105W due to Hubble Ultra Deep Field observations. This is likely due to the 1.083 \micron{} emission line present in this filter \citep{brammer_2014}. \new{In addition, Earthshine can still affect images even after removing images with low Earth limb angles because the Earth limb angle is estimated for each image as an average over the entire exposure.} Remaining outliers will be carefully considered and manually removed before performing Diffuse Light analysis in a future work.

\subsection{Sky-SB SED}

In order to demonstrate how our generated sky-SB measurements can recover the ZL surface brightness and compare our results to the literature, we create a SED of the observed sky-SB at HST wavelengths. \newnew{Figure \ref{fig:sky_vs_wave_pole} shows our sky-SB SED at high ecliptic latitude (within 40\degree{} of either ecliptic pole) where the sky-SB is darkest.}

\new{Systematic errors in estimating the sky-SB at a given position (e.g., fluctuations in DGL, faint stars, etc) are much bigger than the sub-percent statistical errors in sky-SB measurement. For the sake of creating a sky-SB SED, these systematic variations in the sky-SB can be resolved if we only keep one sky-SB measurement per position.} We take the median sky-SB of all observations taken within 10 arcmin of each other, and define each of these 10 arcmin groups as a unique position. Each SKYSURF point in Figure \ref{fig:sky_sed_ecllat_and_sunang} represents the median of all unique positions for every HST filter. The error bars are a combination of the standard error in sky-SB values along different unique positions ($\sigma_\text{spread}$), and the sky-SB error ($\sigma_\text{sky}$):

\begin{equation}
    \sigma_\text{spread} = \frac{\newnew{\sigma_{\text{std}}}}{\sqrt\text{\# of Unique Positions}},
\end{equation}

\begin{equation} \label{eq:errorbar}
    \text{Error Bar} = \sqrt{\sigma_\text{spread}^2 + \sigma_\text{sky, med}^2},
\end{equation}

\noindent \newnew{where $\sigma_{\text{std}}$ is the standard deviation in sky-SB values}, and $\sigma_\text{sky, med}$ is a median of all sky-SB measurement errors (Table \ref{tab:error}). Only filters with at least two unique positions are plotted.

We compare our sky-SB SED in Figure \ref{fig:sky_vs_wave_pole} to several different measurements of the dark sky-SB and estimates of ZL emission. SKYSURF measurements are shown as filled black squares (ACS/WFC), circles (WFC3/UVIS) and triangles (WFC3/IR). The grey dashed line represents the parametric ZL emission model from \cite{aldering_2001} (which is a modification of the \cite{leinert98} model). \new{This is produced} using the \texttt{gunagala} \texttt{sky} module \citep{thomas_2022}, which calculates a ZL model at the ecliptic pole, from which a relative scaling to the appropriate ecliptic latitude is applied. \new{\cite{kawara_2017} measurements of ZL emission taken with the HST's Faint Object Spectrograph are shown as open red circles, and are estimated for an ecliptic latitude of 85\degree{} using Table 2 and Equation 8 from \cite{kawara_2017}.} \new{The open brown diamonds represent the sky-SB measurements used in \citetalias{carleton_2022}, which were estimated using a preliminary sky estimation algorithm.} \cite{giavalisco_2002} present sky-SB measurements scaled to the North Ecliptic Pole, shown as blue plus signs, which include measurements from \cite{leinert98}, the HDF Team, \cite{wright_2001}, and \cite{aldering_2001}.

\newnew{As described in \citetalias{windhorst_2022}, we ignore all resolved galaxies in HST images, while other studies tend to include it during EBL analysis.} Therefore, some sky-SB measurements are not comparable to ours because they include some or all of the flux from resolved galaxies and stars, and thus are higher in value. However, we show them to highlight general consistencies. The \cite{hauser_1998} sky-SB measurement, shown as an open green square, represents the COBE/ DIRBE sky-SB measurement that includes \textit{all} EBL (discrete$+$diffuse). The open orange triangles show dark sky HST WFPC2 measurements from \cite{bernstein_2002}, which excludes EBL coming from stars and galaxies with total magnitudes brighter than AB $\simeq23$ mag (in WFPC2 F555W filter). \new{For comparison, the median SKYSURF exposure is complete to AB $\simeq26$ mag \citepalias{carleton_2022}.} Finally, the purple X's represent sky spectra approximated from Figure 2 of \cite{matsuura_2017} \new{for the North Ecliptic Pole}, which were measured using the Low-resolution Spectrometer on the Cosmic Infrared Background Experiment (CIBER), and also include all EBL.

In summary the shape of our sky-SB measurements agree well with other models and predictions, although some differences are still present as discussed above. At wavelengths shortward of 0.5 \micron{}, our sky-SB measurements in general agree with \cite{kawara_2017}, \cite{bernstein_2002}, and \cite{giavalisco_2002}, which may suggest that the amount of EBL present at UV wavelengths is small.

Between 0.5 \micron{} and 0.8 \micron{}, some SKYSURF measurements do not seem to agree with other measurements. The \cite{bernstein_2002} points are expected to be higher since their measurements include the flux from objects fainter than AB $\simeq23$ mag, where SKYSURF excludes the flux from all resolved objects. The offset between SKYSURF measurements and \cite{kawara_2017} indicates a potential for some Diffuse Light signal between 0.5 \micron{} and 0.8 \micron{}.

SKYSURF points greater than $0.8$ \micron{} agree with other measurements, although the \cite{aldering_2001} ZL model appears to overestimate sky-SB levels in the near-IR. \new{This model is a solar spectrum that has been reddened and scaled, and thus includes many free parameters that could cause the offset seen here (namely the reddening factor).}
The COBE/ DIRBE sky-SB measurement \citep{hauser_1998} is brighter because it includes all EBL, while our sky-SB measurements ignores discrete objects. The CIBER \citep{matsuura_2017} measurements also include all EBL flux. \new{The offset in sky-SB measurements between this work and \citetalias{carleton_2022}, most notably in F125W and F160W, imply that Diffuse Light limits will be lower than in \citetalias{carleton_2022}. We describe this in detail in Section \ref{sec:diffuse_light_limits}.}

The top panel of Figure \ref{fig:sky_sed_ecllat_and_sunang} displays our sky-SB SED for different ecliptic latitude ($b$) bins. \new{We find the shape of the sky-SB SED to be largely similar, following a ZL SED at all ecliptic latitudes. Specifically, using a simple linear least-squares regression \citep{scipy_2020}, the slope of $\log$(sky-SB) versus $\log(\lambda)$ is consistently $\sim$2.2 across all ecliptic latitude bins. The $60\degree<b<80\degree$ bin is an outlier, with a slope of $\sim$2.6. This implies that the sky-SB is dominated by ZL at all ecliptic latitudes.} Following the trends seen in Figures \ref{fig:sky_vs_ecllat_wfc3uvis_1}--\ref{fig:sky_vs_ecllat_wfc3ir}, the darkest sky-SB levels correspond to the highest ecliptic latitudes.

The bottom panel of Figure \ref{fig:sky_sed_ecllat_and_sunang} displays the sky-SB SED for different Sun angle bins. \cite{caddy_2022} and \cite{leinert98} show that Sun angle can influence the brightness of ZL emission, and thus the observed sky-SB. \new{Sun angle can correlate with ecliptic latitude, where high ecliptic latitudes will always have Sun angles $\sim$90\degree. Therefore, we limit our Sun angle bins to have ecliptic latitudes $-30\degree < b < 30\degree$.} We find the shape of our sky-SB SED to depend on Sun angle at wavelengths between 0.9--1.4 \micron. For low Sun angles (grey points), \new{the sky-SB shows a clear peak at $\sim1.4$\micron}. At higher Sun angles, the sky-SB SED flattens between 0.9--1.4 \micron. Although the reason for this change in shape is unknown, it could be caused by scattered light, or indicate that the scattering of solar photons off the IPD is non-isotropic. In other words, it could suggest that the anisotropy of the scattering phase function may change as a function of wavelength. \new{In addition, there may be an additional component to ZL \citep[e.g.,][]{kawara_2017, korngut_2022} that interacts with photons with wavelengths between 1.1--1.6 \micron{} differently.}

\subsection{Sky-SB rms}\label{sec:skyrms}

In Figure \ref{fig:rms_vs_wave}, we show the median measured sky-SB rms for each filter. WFC3/IR shows lower measured sky-SB rms values likely due to the larger pixel size compared to WFC3/UVIS and ACS/WFC (see pixel sizes described in Appendix \ref{app:converting_to_flux_units}).

\newnew{We note that the F850LP has consistently more scatter when compared to other filters, which is most clearly seen in Figure \ref{fig:rms_vs_wave}. This is likely driven by the small number of WFC3/UVIS F850LP observations, most of which are at ecliptic latitudes $<30$\degree{} (Figure \ref{fig:sky_vs_ecllat_wfc3uvis_2}), where the sky-SB and the sky-SB rms is higher.}

\section{Updated Diffuse Light Limits} \label{sec:diffuse_light_limits}

\new{The main goal of SKYSURF is to provide Diffuse Light (DL) measurements using HST's vast archive. \citetalias{carleton_2022} presents DL upper limits using preliminary sky-SB measurement algorithms that are meant to be conservative (and computationally faster), and provide initial DL limits using three pilot near-infrared filters: F125W, F140W and F160W. With the improved sky-SB measurement algorithms in this work, we update the DL limits presented in \citetalias{carleton_2022} for the F125W, F140W and F160W filters. DL measurements for all filters, using improved SKYSURF ZL models, will be presented in future work.}

The DL upper limits from \citetalias{carleton_2022} are estimated using the Lowest Fitted Sky (LFS) method, where they fit a $sech$ curve to the darkest thermal-dark corrected sky-SB values measured for SKYSURF. They fit a similar curve to the \citet{kelsall98} ZL emission predictions and estimate a DL signal by comparing the two curves. DGL and unresolved EBL still present in HST images is also subtracted during this process. \new{\citetalias{carleton_2022} DL limits are presented as upper limits due to uncertainties in thermal dark, which will be updated in Carleton et al. (in prep).}

\newnew{Revised DL upper limits are shown in Table \ref{tab:diffuse_light}.} We estimate these limits by calculating the 3$\sigma$-clipped median difference in sky-SB values between the sky-SB measurements in this work and the sky-SB measurements used in \citetalias{carleton_2022}. In Appendix \ref{app:diffuse_light_confirm}, we confirm that a 3$\sigma$-clipped median difference is a good representation of the darkest sky-SB measurements used in \citetalias{carleton_2022}. The \citet{kelsall98} Zodiacal model predictions do not change between this work and \citetalias{carleton_2022}. \newnew{The improved algorithms in this work result in sky-SB measurements that are $\sim1-2\%$ lower than the F125W, F140W, and F160W measurements in \citetalias{carleton_2022}. These updated sky-SB measurements result in DL upper limits that are typically 20--30\% lower than the conservative limits from \citetalias{carleton_2022}.}

\new{DL limits for the entire HST wavelength range will be estimated in future papers, using a ZL model constrained by the sky-SB measurements in this paper, and supplemented by star and galaxy counts using the SKYSURF database.}

\begin{table*}[t]
    \centering
    \begin{tabular}{|c|c|c|c|}
    \hline
    Filter & F125W & F140W & F160W \\
    \hline
    \citetalias{carleton_2022} DL Limit [\MJysr] & $\lesssim$0.012 & $\lesssim$0.018 & $\lesssim$0.015\\
    \hspace{110pt} $[$\nWmsr$]$           &    $\lesssim$29 &   $\lesssim$40  & $\lesssim$29\\
    \hline
    Ratio: Sky-SB / \citetalias{carleton_2022} Sky-SB                 & 0.983  & 0.986  & 0.988 \\
    Difference: Sky-SB $-$ \citetalias{carleton_2022} Sky-SB [\MJysr] & -0.003 & -0.003 & -0.002 \\
    \hline
    \textbf{This Work: DL Limit [\MJysr]} & \textbf{$\lesssim$0.009} & \textbf{$\lesssim$0.015} & \textbf{$\lesssim$0.013} \\
    \hspace{119pt} \textbf{$[$\nWmsr$]$}           & \textbf{$\lesssim$22}             & \textbf{$\lesssim$32}             & \textbf{$\lesssim$25} \\
    \hline
    \end{tabular}
    \caption{Updated SKYSURF Diffuse Light limits. We compare sky-SB measurements for F125W, F140W, and F160W from \citetalias{carleton_2022} to this work. The first row lists the DL limits from \citetalias{carleton_2022}. The second row shows the same results in units of \nWmsr. The third row shows the median ratio between sky-SB measurements from this work and those from \citetalias{carleton_2022}. The fourth row shows the difference in these sky-SB values, in units of \MJysr. We subtract this difference from the SKYSURF-2 DL limit in the first row to estimate a DL limit for this work, shown in the fifth row. The last row shows the same DL limits from the fifth row in units of \nWmsr. We adopt the same error from \citetalias{carleton_2022}: 0.005 \MJysr ($\sim10$ \nWmsr).}
    \label{tab:diffuse_light}
\end{table*}

\section{Public Data Products} \label{sec:data}

We provide several data products on the official SKYSURF website\footnote{\url{http://skysurf.asu.edu}}. Relevant to this paper are: 1) sky-SB measurements for all SKYSURF images, and 2) FITS files containing the sky sub-regions used for our algorithm. In addition, the SKYSURF website presents the Pro-med algorithm used for sky-SB estimation in this work.

\subsection{Sky-SB Data Tables}

For every image, we provide a SKYSURF sky-SB measurement in the native units of the \flt{}/\flc{} files, as well as in calibrated flux units of \MJysr. The process for converting our images to units of flux density is described in Appendix \ref{app:converting_to_flux_units}. Estimates of the thermal dark levels will be presented in Carleton et al. (in prep) and are included in the public files for all WFC3/IR measurements. We provide sky-SB measurements with and without thermal dark corrections.

We include an uncertainty for each sky-SB measurement using Table \ref{tab:error}. The error is calculated as following:

\begin{equation} \label{eq:sigma_sky}
    \sigma_{\text{sky}} = \sqrt{\sigma_{\text{add}}^2 + (\sigma_{\text{mult}}\times \skyim)^2},
\end{equation}

\noindent where $\sigma_{\text{sky}}$ is the total sky-SB error in units of electrons or electrons per second, $\sigma_{\text{add}}$ is the additive error in units of electrons or electrons per second, $\sigma_{\text{mult}}$ is the multiplicative error in units of percent, and $\skyim$ is the measured sky-SB in units of electrons or electrons per second. The sky-SB error is also presented in units of \MJysr.

Every sky-SB measurement has a corresponding flag that designates images with too many bad sub-regions, too high of a sky-SB rms, images marked during the manual inspections, located within galactic plane, located close to the Earth's limb, located at high Sun altitude's, contains a large common object, or has too many pixels affected by persistence. We refer to corresponding documentation on the public files for more information.

\subsection{Subregion FITS Files}

We provide FITS files containing the sub-regions created during our algorithm described in Section \ref{sec:methods}. Each FITS file, which we refer to as a SUB file, has a single sky value associated with each sub-region where each sub-region takes up a single pixel. For example, a $1014\times1014$ pixel WFC3-IR image that is divided into $26\times26$ pixel sub-regions will have a corresponding SUB file that is $26\times26$ pixels in size.

They contain one primary header and two data extensions. The primary header is copied from the original \flt{}/\flc{} image, where the extension name (EXTNAME) is changed to SUB\_SKY. The two data extensions are labeled `SKY' and `RMS' and contain the sky-SB and sky-SB rms sub-region data. In the SKY and RMS extensions only, WCS keywords are updated so that the SUB files map onto the true sky to within 0.3 arcseconds.

\section{Conclusion}

We present sky-SB algorithms and measurements for project SKYSURF, an HST archival program with the end goal of constraining an all-sky DL signal. \new{SKYSURF is the first study of the sky-SB with HST at this scale and encompassing HST’s entire wavelength range.} The SKYSURF database includes more than 140,000 HST images, spanning 0.2--1.6 \micron{}.

We utilize the Pro-med algorithm to measure the sky-SB (Figure \ref{fig:methods_flowchart}) for all images in our database. As a function of ecliptic latitude, Figures \ref{fig:sky_vs_ecllat_wfc3uvis_1} -- \ref{fig:sky_sed_ecllat_and_sunang} follow expected trends by peaking near the ecliptic plane. There are almost no outliers \new{falling outside the expected trend}, highlighting the success of our ability to filter our unreliable sky-SB measurements. The sky-SB SED shows that our measurements generally agree well with other measurements. \newnew{The overall shape of our sky-SB SED for low Sun angles does not match that of higher Sun angles, where the low Sun angle sky-SB SED shows a peak between $\sim1.1-1.6$ \micron.} The cause for this remains unknown, yet we propose it may be due to the anisotropy of the scattering phase function of interplanetary dust, or due to an additional ZL component in the inner Solar System.   

As shown in Table \ref{tab:diffuse_light}, we estimate DL limits based on the methods of \citetalias{carleton_2022} for F125W, F140W, and F160W. The DL limits in \citetalias{carleton_2022} are conservative, and the DL limits in this work are lower those in \citetalias{carleton_2022}, ranging from 0.009 \MJysr{} to 0.015 \MJysr. The DL limits presented here are still designed to be conservative, and measurements of DL using an updated ZL model for the entire HST wavelength range will be provided in future papers. Overall, these estimates provide the most stringent all-sky constraints in this wavelength range, and show a significant DL component of unknown origin. 

We hope that these sky-SB measurements will not only benefit Project SKYSURF, but will \new{provide valuable data and methods for future programs}. SKYSURF data products are released to the public (Section \ref{sec:data}), including tables of all SKYSURF sky-SB measurements, as well as FITS files that show the sub-regions used during our sky-SB estimation. \new{We also make our sky-SB algorithms available to the public. Aside from using our methods to study the sky-SB itself,} reliable sky-SB algorithms are crucial for reliable photometry for low surface-brightness studies, where it is imperative that signal from real objects does not contaminate the measured sky-SB.

\new{SKYSURF's large database gives us the unique ability to independently and consistently derive galaxy counts (and therefore create a SKYSURF EBL model), as well as constrain ZL emission at HST wavelengths. We can compare our sky-SB measurements to a SKYSURF EBL model based on galaxy counts and a SKYSURF ZL emission model for a final estimate of DL. Future work includes generating source counts using the entire SKYSURF database, updated Thermal Dark signals, constraining a ZL model utilizing SKYSURF data, and measuring an EBL signal using these results.}

\section*{Acknowledgements}

All of the data presented in this paper were obtained from the Mikulski Archive for Space Telescopes (MAST). This project is based on observations made with the NASA/ESA Hubble Space Telescope and obtained from the Hubble Legacy Archive, which is a collaboration between the Space Telescope Science Institute (STScI/NASA), the Space Telescope European Coordinating Facility (ST-ECF/ESA), and the Canadian Astronomy Data Centre (CADC/NRC/CSA). Some image simulations were based on observations taken by the 3D-HST Treasury Program (GO 12177 and 12328) with the NASA/ESA HST, which is operated by the Association of Universities for Research in Astronomy, Inc., under NASA contract NAS5-26555.

We acknowledge support for HST programs AR-09955 and AR-15810 provided by NASA through grants from the Space Telescope Science Institute, which is operated by the Association of Universities for Research in Astronomy, Incorporated, under NASA contract NAS5-26555. Work by R.G.A. was supported by NASA under award number 80GSFC21M0002.

We especially thank the entire SKYSURF team for their contributions to making this project successful. We also thank the thoughtful for reviewer for the beneficial comments that helped clarify this work. 

We also acknowledge the indigenous peoples of Arizona, including the Akimel O’odham (Pima) and Pee Posh (Maricopa) Indian Communities, whose care and keeping of the land has enabled us to be at ASU’s Tempe campus in the Salt River Valley, where this work was conducted.

\facility{Hubble Space Telescope Mikulski Archive \url{https://archive.stsci.edu}; Hubble Legacy Archive (HLA) \url{https://hla.stsci.edu}; Hubble Legacy Catalog (HLC) \url{https://archive.stsci.edu/hst/hsc/}}

\software{Astropy \citep{astropy:2013, astropy:2018, astropy:2022}; ProFound \citep{robotham_2018}; GalSim \citep{rowe15}}

\bibliography{main_skysurf_methods}{}

\begin{thebibliography}{}
\expandafter\ifx\csname natexlab\endcsname\relax\def\natexlab#1{#1}\fi
\providecommand{\url}[1]{\href{#1}{#1}}
\providecommand{\dodoi}[1]{doi:~\href{http://doi.org/#1}{\nolinkurl{#1}}}
\providecommand{\doeprint}[1]{\href{http://ascl.net/#1}{\nolinkurl{http://ascl.net/#1}}}
\providecommand{\doarXiv}[1]{\href{https://arxiv.org/abs/#1}{\nolinkurl{https://arxiv.org/abs/#1}}}

\bibitem[{{Aldering}(2001)}]{aldering_2001}
{Aldering}, G. 2001, LBNL report, LBNL-51157, 1,
  \dodoi{http://www-supernova.lbl.gov/~aldering/}

\bibitem[{{Anand} {et~al.}(2022){Anand}, {Grogin}, \& {Anderson}}]{anand_2022}
{Anand}, G., {Grogin}, N., \& {Anderson}, J. 2022, {Revisiting ACS/WFC Sky
  Backgrounds}, Instrument Science Report ACS 2022-1

\bibitem[{Andrews {et~al.}(2017)Andrews, Driver, Davies, Lagos, \&
  Robotham}]{andrews_2017}
Andrews, S.~K., Driver, S.~P., Davies, L.~J., Lagos, C. d.~P., \& Robotham, A.
  S.~G. 2017, arXiv, 474, 898–916, \dodoi{10.1093/mnras/stx2843}

\bibitem[{{Ashcraft} {et~al.}(2018){Ashcraft}, {Windhorst}, {Jansen}, {Cohen},
  {Grazian}, {Paris}, {Fontana}, {Giallongo}, {Speziali}, {Testa}, {Boutsia},
  {O'Connell}, {Rutkowski}, {Ryan}, {Scarlata}, \& {Weiner}}]{ashcraft_2018a}
{Ashcraft}, T.~A., {Windhorst}, R.~A., {Jansen}, R.~A., {et~al.} 2018, \pasp,
  130, 064102, \dodoi{10.1088/1538-3873/aab542}

\bibitem[{{Astropy Collaboration} {et~al.}(2013){Astropy Collaboration},
  {Robitaille}, {Tollerud}, {Greenfield}, {Droettboom}, {Bray}, {Aldcroft},
  {Davis}, {Ginsburg}, {Price-Whelan}, {Kerzendorf}, {Conley}, {Crighton},
  {Barbary}, {Muna}, {Ferguson}, {Grollier}, {Parikh}, {Nair}, {Unther},
  {Deil}, {Woillez}, {Conseil}, {Kramer}, {Turner}, {Singer}, {Fox}, {Weaver},
  {Zabalza}, {Edwards}, {Azalee Bostroem}, {Burke}, {Casey}, {Crawford},
  {Dencheva}, {Ely}, {Jenness}, {Labrie}, {Lim}, {Pierfederici}, {Pontzen},
  {Ptak}, {Refsdal}, {Servillat}, \& {Streicher}}]{astropy:2013}
{Astropy Collaboration}, {Robitaille}, T.~P., {Tollerud}, E.~J., {et~al.} 2013,
  \aap, 558, A33, \dodoi{10.1051/0004-6361/201322068}

\bibitem[{{Astropy Collaboration} {et~al.}(2018){Astropy Collaboration},
  {Price-Whelan}, {Sip{\H{o}}cz}, {G{\"u}nther}, {Lim}, {Crawford}, {Conseil},
  {Shupe}, {Craig}, {Dencheva}, {Ginsburg}, {Vand erPlas}, {Bradley},
  {P{\'e}rez-Su{\'a}rez}, {de Val-Borro}, {Aldcroft}, {Cruz}, {Robitaille},
  {Tollerud}, {Ardelean}, {Babej}, {Bach}, {Bachetti}, {Bakanov}, {Bamford},
  {Barentsen}, {Barmby}, {Baumbach}, {Berry}, {Biscani}, {Boquien}, {Bostroem},
  {Bouma}, {Brammer}, {Bray}, {Breytenbach}, {Buddelmeijer}, {Burke},
  {Calderone}, {Cano Rodr{\'\i}guez}, {Cara}, {Cardoso}, {Cheedella}, {Copin},
  {Corrales}, {Crichton}, {D'Avella}, {Deil}, {Depagne}, {Dietrich}, {Donath},
  {Droettboom}, {Earl}, {Erben}, {Fabbro}, {Ferreira}, {Finethy}, {Fox},
  {Garrison}, {Gibbons}, {Goldstein}, {Gommers}, {Greco}, {Greenfield},
  {Groener}, {Grollier}, {Hagen}, {Hirst}, {Homeier}, {Horton}, {Hosseinzadeh},
  {Hu}, {Hunkeler}, {Ivezi{\'c}}, {Jain}, {Jenness}, {Kanarek}, {Kendrew},
  {Kern}, {Kerzendorf}, {Khvalko}, {King}, {Kirkby}, {Kulkarni}, {Kumar},
  {Lee}, {Lenz}, {Littlefair}, {Ma}, {Macleod}, {Mastropietro}, {McCully},
  {Montagnac}, {Morris}, {Mueller}, {Mumford}, {Muna}, {Murphy}, {Nelson},
  {Nguyen}, {Ninan}, {N{\"o}the}, {Ogaz}, {Oh}, {Parejko}, {Parley}, {Pascual},
  {Patil}, {Patil}, {Plunkett}, {Prochaska}, {Rastogi}, {Reddy Janga},
  {Sabater}, {Sakurikar}, {Seifert}, {Sherbert}, {Sherwood-Taylor}, {Shih},
  {Sick}, {Silbiger}, {Singanamalla}, {Singer}, {Sladen}, {Sooley},
  {Sornarajah}, {Streicher}, {Teuben}, {Thomas}, {Tremblay}, {Turner},
  {Terr{\'o}n}, {van Kerkwijk}, {de la Vega}, {Watkins}, {Weaver}, {Whitmore},
  {Woillez}, {Zabalza}, \& {Astropy Contributors}}]{astropy:2018}
{Astropy Collaboration}, {Price-Whelan}, A.~M., {Sip{\H{o}}cz}, B.~M., {et~al.}
  2018, \aj, 156, 123, \dodoi{10.3847/1538-3881/aabc4f}

\bibitem[{{Astropy Collaboration} {et~al.}(2022){Astropy Collaboration},
  {Price-Whelan}, {Lim}, {Earl}, {Starkman}, {Bradley}, {Shupe}, {Patil},
  {Corrales}, {Brasseur}, {N{"o}the}, {Donath}, {Tollerud}, {Morris},
  {Ginsburg}, {Vaher}, {Weaver}, {Tocknell}, {Jamieson}, {van Kerkwijk},
  {Robitaille}, {Merry}, {Bachetti}, {G{"u}nther}, {Aldcroft},
  {Alvarado-Montes}, {Archibald}, {B{'o}di}, {Bapat}, {Barentsen}, {Baz{'a}n},
  {Biswas}, {Boquien}, {Burke}, {Cara}, {Cara}, {Conroy}, {Conseil}, {Craig},
  {Cross}, {Cruz}, {D'Eugenio}, {Dencheva}, {Devillepoix}, {Dietrich},
  {Eigenbrot}, {Erben}, {Ferreira}, {Foreman-Mackey}, {Fox}, {Freij}, {Garg},
  {Geda}, {Glattly}, {Gondhalekar}, {Gordon}, {Grant}, {Greenfield}, {Groener},
  {Guest}, {Gurovich}, {Handberg}, {Hart}, {Hatfield-Dodds}, {Homeier},
  {Hosseinzadeh}, {Jenness}, {Jones}, {Joseph}, {Kalmbach}, {Karamehmetoglu},
  {Ka{l}uszy{'n}ski}, {Kelley}, {Kern}, {Kerzendorf}, {Koch}, {Kulumani},
  {Lee}, {Ly}, {Ma}, {MacBride}, {Maljaars}, {Muna}, {Murphy}, {Norman},
  {O'Steen}, {Oman}, {Pacifici}, {Pascual}, {Pascual-Granado}, {Patil},
  {Perren}, {Pickering}, {Rastogi}, {Roulston}, {Ryan}, {Rykoff}, {Sabater},
  {Sakurikar}, {Salgado}, {Sanghi}, {Saunders}, {Savchenko}, {Schwardt},
  {Seifert-Eckert}, {Shih}, {Jain}, {Shukla}, {Sick}, {Simpson},
  {Singanamalla}, {Singer}, {Singhal}, {Sinha}, {Sip{H{o}}cz}, {Spitler},
  {Stansby}, {Streicher}, {{{S}}umak}, {Swinbank}, {Taranu}, {Tewary},
  {Tremblay}, {Val-Borro}, {Van Kooten}, {Vasovi{'c}}, {Verma}, {de Miranda
  Cardoso}, {Williams}, {Wilson}, {Winkel}, {Wood-Vasey}, {Xue}, {Yoachim},
  {Zhang}, {Zonca}, \& {Astropy Project Contributors}}]{astropy:2022}
{Astropy Collaboration}, {Price-Whelan}, A.~M., {Lim}, P.~L., {et~al.} 2022,
  apj, 935, 167, \dodoi{10.3847/1538-4357/ac7c74}

\bibitem[{{Bernstein} {et~al.}(1995){Bernstein}, {Nichol}, {Tyson}, {Ulmer}, \&
  {Wittman}}]{bernstein_1995}
{Bernstein}, G.~M., {Nichol}, R.~C., {Tyson}, J.~A., {Ulmer}, M.~P., \&
  {Wittman}, D. 1995, \aj, 110, 1507, \dodoi{10.1086/117624}

\bibitem[{Bernstein(2007)}]{bernstein_2007}
Bernstein, R.~A. 2007, The Astrophysical Journal, 666, 663–673,
  \dodoi{10.1086/519824}

\bibitem[{Bernstein {et~al.}(2002)Bernstein, Freedman, \&
  Madore}]{bernstein_2002}
Bernstein, R.~A., Freedman, W.~L., \& Madore, B.~F. 2002, The Astrophysical
  Journal, 571, 56–84, \dodoi{10.1086/339422}

\bibitem[{{Bertin} \& {Arnouts}(1996)}]{bertin96}
{Bertin}, E., \& {Arnouts}, S. 1996, \aaps, 117, 393,
  \dodoi{10.1051/aas:1996164}

\bibitem[{{Bickel} \& {Fruehwirth}(2005)}]{bickel05}
{Bickel}, D.~R., \& {Fruehwirth}, R. 2005, arXiv Mathematics e-prints,
  math/0505419.
\newblock \doarXiv{math/0505419}

\bibitem[{{Biretta} \& {Baggett}(2013)}]{biretta_2013}
{Biretta}, J., \& {Baggett}, S. 2013, {WFC3 Post-Flash Calibration}, Space
  Telescope WFC Instrument Science Report

\bibitem[{{Bohlin} {et~al.}(2020){Bohlin}, {Ryon}, \& {Anderson}}]{bohlin_2020}
{Bohlin}, R.~C., {Ryon}, J.~E., \& {Anderson}, J. 2020, {Update of the
  Photometric Calibration of the ACS CCD Cameras}, Instrument Science Report
  ACS 2020-8

\bibitem[{{Borlaff} {et~al.}(2019){Borlaff}, {Trujillo}, {Rom{\'a}n},
  {Beckman}, {Eliche-Moral}, {Infante-S{\'a}inz}, {Lumbreras-Calle}, {de
  Almagro}, {G{\'o}mez-Guijarro}, {Cebri{\'a}n}, {Dorta}, {Cardiel},
  {Akhlaghi}, \& {Mart{\'\i}nez-Lombilla}}]{borlaff_2019}
{Borlaff}, A., {Trujillo}, I., {Rom{\'a}n}, J., {et~al.} 2019, \aap, 621, A133,
  \dodoi{10.1051/0004-6361/201834312}

\bibitem[{Bourque \& Baggett(2016)}]{bourque_2016}
Bourque, M., \& Baggett, S. 2016

\bibitem[{Bradley {et~al.}(2020)Bradley, Sip{\H o}cz, Robitaille, Tollerud,
  Vin{\'{\i}}cius, Deil, Barbary, Wilson, Busko, G{\"u}nther, Cara, Conseil,
  Bostroem, Droettboom, Bray, Bratholm, Lim, Barentsen, Craig, Pascual, Perren,
  Greco, Donath, de~Val-Borro, Kerzendorf, Bach, Weaver, D'Eugenio, Souchereau,
  \& Ferreira}]{bradley20}
Bradley, L., Sip{\H o}cz, B., Robitaille, T., {et~al.} 2020, astropy/photutils:
  1.0.0, 1.0.0,  Zenodo, \dodoi{10.5281/zenodo.4044744}

\bibitem[{{Brammer} {et~al.}(2014){Brammer}, {Pirzkal}, {McCullough}, \&
  {MacKenty}}]{brammer_2014}
{Brammer}, G., {Pirzkal}, N., {McCullough}, P., \& {MacKenty}, J. 2014,
  {Time-varying Excess Earth-glow Backgrounds in the WFC3/IR Channel},
  Instrument Science Report WFC3 2014-03, 14 pages

\bibitem[{{Caddy} {et~al.}(2022){Caddy}, {Spitler}, \& {Ellis}}]{caddy_2022}
{Caddy}, S.~E., {Spitler}, L.~R., \& {Ellis}, S.~C. 2022, \aj, 164, 52,
  \dodoi{10.3847/1538-3881/ac76c2}

\bibitem[{Calamida {et~al.}(2022)Calamida, Bajaj, Mack, Marinelli, Medina,
  Pidgeon, Kozhurina-Platais, Shanahan, \& Som}]{calamida_2022}
Calamida, A., Bajaj, V., Mack, J., {et~al.} 2022, arXiv

\bibitem[{Cambresy {et~al.}(2001)Cambresy, Reach, Beichman, \&
  Jarrett}]{cambresy_2001}
Cambresy, L., Reach, W.~T., Beichman, C.~A., \& Jarrett, T.~H. 2001, The
  Astrophysical Journal, 555, 563–571, \dodoi{10.1086/321470}

\bibitem[{{Carleton} {et~al.}(2022){Carleton}, {Windhorst}, {O'Brien}, {Cohen},
  {Carter}, {Jansen}, {Tompkins}, {Arendt}, {Caddy}, {Grogin}, {Kenyon},
  {Koekemoer}, {MacKenty}, {Casertano}, {Davies}, {Driver}, {Dwek},
  {Kashlinsky}, {Miles}, {Pawnikar}, {Pirzkal}, {Robotham}, {Ryan}, {Abate},
  {Andras-Letanovszky}, {Berkheimer}, {Goisman}, {Henningsen}, {Kramer},
  {Rogers}, \& {Swirbul}}]{carleton_2022}
{Carleton}, T., {Windhorst}, R.~A., {O'Brien}, R., {et~al.} 2022, arXiv
  e-prints, arXiv:2205.06347.
\newblock \doarXiv{2205.06347}

\bibitem[{Cohen {et~al.}(2020)Cohen, Grogin, \& Bellini}]{cohen_2020}
Cohen, Y., Grogin, N.~A., \& Bellini, A. 2020, Space Telescope ACS Instrument
  Science Report

\bibitem[{Conselice {et~al.}(2016)Conselice, Wilkinson, Duncan, \&
  Mortlock}]{conselice_2016}
Conselice, C.~J., Wilkinson, A., Duncan, K., \& Mortlock, A. 2016, The
  Astrophysical Journal, 830, 83, \dodoi{10.3847/0004-637x/830/2/83}

\bibitem[{{Cooray} {et~al.}(2004){Cooray}, {Bock}, {Keatin}, {Lange}, \&
  {Matsumoto}}]{cooray_2004}
{Cooray}, A., {Bock}, J.~J., {Keatin}, B., {Lange}, A.~E., \& {Matsumoto}, T.
  2004, \apj, 606, 611, \dodoi{10.1086/383137}

\bibitem[{{de Vaucouleurs} {et~al.}(1991){de Vaucouleurs}, {de Vaucouleurs},
  {Corwin}, {Buta}, {Paturel}, \& {Fouque}}]{de_Vaucouleurs_1991}
{de Vaucouleurs}, G., {de Vaucouleurs}, A., {Corwin}, Herold~G., J., {et~al.}
  1991, {Third Reference Catalogue of Bright Galaxies}

\bibitem[{{Dole, H.} {et~al.}(2006){Dole, H.}, {Lagache, G.}, {Puget, J.-L.},
  {Caputi, K. I.}, {Fern\'andez-Conde, N.}, {Le Floc'h, E.}, {Papovich, C.},
  {P\'erez-Gonz\'alez, P. G.}, {Rieke, G. H.}, \& {Blaylock, M.}}]{dole_2006}
{Dole, H.}, {Lagache, G.}, {Puget, J.-L.}, {et~al.} 2006, A\&A, 451, 417,
  \dodoi{10.1051/0004-6361:20054446}

\bibitem[{Domínguez {et~al.}(2011)Domínguez, Primack, Rosario, Prada,
  Gilmore, Faber, Koo, Somerville, Pérez‐Torres, Pérez‐González, Huang,
  Davis, Guhathakurta, Barmby, Conselice, Lozano, Newman, \&
  Cooper}]{dominguez_2011}
Domínguez, A., Primack, J.~R., Rosario, D.~J., {et~al.} 2011, Monthly Notices
  of the Royal Astronomical Society, 410, 2556–2578,
  \dodoi{10.1111/j.1365-2966.2010.17631.x}

\bibitem[{{Dressel}(2021)}]{wfc3_ihb}
{Dressel}, L. 2021, in WFC3 Instrument Handbook for Cycle 29 v. 13, Vol.~13, 13

\bibitem[{Driver(2021)}]{driver_2021}
Driver, S.~P. 2021, arXiv

\bibitem[{Driver {et~al.}(2016)Driver, Andrews, Davies, Robotham, Wright,
  Windhorst, Cohen, Emig, Jansen, \& Dunne}]{driver16}
Driver, S.~P., Andrews, S.~K., Davies, L.~J., {et~al.} 2016, The Astrophysical
  Journal, 827, 108, \dodoi{10.3847/0004-637x/827/2/108}

\bibitem[{Dwek \& Arendt(1999)}]{dwek_1999}
Dwek, E., \& Arendt, R.~G. 1999, AIP Conference Proceedings, 470, 354–358,
  \dodoi{10.1063/1.58621}

\bibitem[{{Gennaro}(2018)}]{wfc3_dhb}
{Gennaro}, M. 2018, {``WFC3 Data Handbook'', Version 4.0}, (Baltimore: STScI)

\bibitem[{{Giavalisco}(2004)}]{giavalisco_2004}
{Giavalisco}, M. 2004, {Cross-Talk in the ACS WFC Detectors. II: Using GAIN=2
  to Minimize the Effect}, Instrument Science Report ACS 2004-13

\bibitem[{{Giavalisco} {et~al.}(2002){Giavalisco}, {Sahu}, \&
  {Bohlin}}]{giavalisco_2002}
{Giavalisco}, M., {Sahu}, K., \& {Bohlin}, R.~C. 2002, {New Estimates of the
  Sky Background for the HST Exposure Time Calculator}, Instrument Science
  Report WFC3 2002-12, 9 pages

\bibitem[{{Gilhuly} {et~al.}(2022){Gilhuly}, {Merritt}, {Abraham}, {Danieli},
  {Lokhorst}, {Liu}, {van Dokkum}, {Conroy}, \& {Greco}}]{gilhuly_2022}
{Gilhuly}, C., {Merritt}, A., {Abraham}, R., {et~al.} 2022, \apj, 932, 44,
  \dodoi{10.3847/1538-4357/ac6750}

\bibitem[{Gong {et~al.}(2016)Gong, Cooray, Mitchell-Wynne, Chen, Zemcov, \&
  Smidt}]{gong_2016}
Gong, Y., Cooray, A., Mitchell-Wynne, K., {et~al.} 2016, The Astrophysical
  Journal, 825, 104, \dodoi{10.3847/0004-637x/825/2/104}

\bibitem[{Hauser {et~al.}(1998)Hauser, Arendt, Kelsall, Dwek, Odegard, Weiland,
  Freudenreich, Reach, Silverberg, Moseley, Pei, Lubin, Mather, Shafer, Smoot,
  Weiss, Wilkinson, \& Wright}]{hauser_1998}
Hauser, M.~G., Arendt, R.~G., Kelsall, T., {et~al.} 1998, The Astrophysical
  Journal, 508, 25–43, \dodoi{10.1086/306379}

\bibitem[{Hill {et~al.}(2018)Hill, Masui, \& Scott}]{hill_2018}
Hill, R., Masui, K.~W., \& Scott, D. 2018, Applied spectroscopy, 72, 663–688,
  \dodoi{10.1177/0003702818767133}

\bibitem[{{Kashlinsky} {et~al.}(2004){Kashlinsky}, {Arendt}, {Gardner},
  {Mather}, \& {Moseley}}]{kashlinsky_2004}
{Kashlinsky}, A., {Arendt}, R., {Gardner}, J.~P., {Mather}, J.~C., \&
  {Moseley}, S.~H. 2004, \apj, 608, 1, \dodoi{10.1086/386365}

\bibitem[{Kawara {et~al.}(2017)Kawara, Matsuoka, Sano, Brandt, Sameshima,
  Tsumura, Oyabu, \& Ienaka}]{kawara_2017}
Kawara, K., Matsuoka, Y., Sano, K., {et~al.} 2017, Publications of the
  Astronomical Society of Japan, 69, \dodoi{10.1093/pasj/psx003}

\bibitem[{Kelsall {et~al.}(1998)Kelsall, Weiland, Franz, Reach, Arendt, Dwek,
  Freudenreich, Hauser, Moseley, Odegard, Silverberg, \& Wright}]{kelsall98}
Kelsall, T., Weiland, J.~L., Franz, B.~A., {et~al.} 1998, The Astrophysical
  Journal, 508, 44–73, \dodoi{10.1086/306380}

\bibitem[{Korngut {et~al.}(2021)Korngut, Kim, Arai, Bangale, Bock, Cooray,
  Cheng, Feder, Hristov, Lanz, Levenson, Matsumoto, Matsuura, Nguyen, Sano,
  Tsumura, \& Zemcov}]{korngut_2021}
Korngut, P., Kim, M.~G., Arai, T., {et~al.} 2021, arXiv

\bibitem[{{Korngut} {et~al.}(2022){Korngut}, {Kim}, {Arai}, {Bangale}, {Bock},
  {Cooray}, {Cheng}, {Feder}, {Hristov}, {Lanz}, {Lee}, {Levenson},
  {Matsumoto}, {Matsuura}, {Nguyen}, {Sano}, {Tsumura}, \&
  {Zemcov}}]{korngut_2022}
{Korngut}, P.~M., {Kim}, M.~G., {Arai}, T., {et~al.} 2022, \apj, 926, 133,
  \dodoi{10.3847/1538-4357/ac44ff}

\bibitem[{Koushan {et~al.}(2021)Koushan, Driver, Bellstedt, Davies, Robotham,
  Lagos, Hashemizadeh, Obreschkow, Thorne, Bremer, Holwerda, Hopkins, Jarvis,
  Siudek, \& Windhorst}]{koushan_2021}
Koushan, S., Driver, S.~P., Bellstedt, S., {et~al.} 2021, arXiv, 503,
  2033–2052, \dodoi{10.1093/mnras/stab540}

\bibitem[{{Lauer} {et~al.}(2021){Lauer}, {Postman}, {Weaver}, {Spencer},
  {Stern}, {Buie}, {Durda}, {Lisse}, {Poppe}, {Binzel}, {Britt}, {Buratti},
  {Cheng}, {Grundy}, {Hor{\'a}nyi}, {Kavelaars}, {Linscott}, {McKinnon},
  {Moore}, {N{\'u}{\~n}ez}, {Olkin}, {Parker}, {Porter}, {Reuter}, {Robbins},
  {Schenk}, {Showalter}, {Singer}, {Verbiscer}, \& {Young}}]{lauer_2021}
{Lauer}, T.~R., {Postman}, M., {Weaver}, H.~A., {et~al.} 2021, \apj, 906, 77,
  \dodoi{10.3847/1538-4357/abc881}

\bibitem[{Lauer {et~al.}(2022)Lauer, Postman, Spencer, Weaver, Stern,
  Gladstone, Binzel, Britt, Buie, Buratti, Cheng, Grundy, Horányi, Kavelaars,
  Linscott, Lisse, McKinnon, McNutt, Moore, Núñez, Olkin, Parker, Porter,
  Reuter, Robbins, Schenk, Showalter, Singer, Verbiscer, \& Young}]{lauer_2022}
Lauer, T.~R., Postman, M., Spencer, J.~R., {et~al.} 2022, The Astrophysical
  Journal Letters, 927, L8, \dodoi{10.3847/2041-8213/ac573d}

\bibitem[{Leinert {et~al.}(1998)Leinert, Bowyer, Haikala, Hanner, Hauser,
  Levasseur-Regourd, Mann, Mattila, Reach, Schlosser, Staude, Toller, Weiland,
  Weinberg, \& Witt}]{leinert98}
Leinert, C., Bowyer, S., Haikala, L.~K., {et~al.} 1998, Astronomy and
  Astrophysics Supplement Series, 127, 1–99, \dodoi{10.1051/aas:1998105}

\bibitem[{{Li} {et~al.}(2022){Li}, {Huang}, {Leauthaud}, {Moustakas},
  {Danieli}, {Greene}, {Abraham}, {Ardila}, {Kado-Fong}, {Lokhorst}, {Lupton},
  \& {Price}}]{li_2022}
{Li}, J., {Huang}, S., {Leauthaud}, A., {et~al.} 2022, \mnras, 515, 5335,
  \dodoi{10.1093/mnras/stac2121}

\bibitem[{Lucas(2021)}]{acs_dhb}
Lucas, R.~A., e.~a. 2021

\bibitem[{Mack {et~al.}(2016)Mack, Dahlen, Sabbi, \& Bowers}]{mack_2016}
Mack, J., Dahlen, T., Sabbi, E., \& Bowers, . A.~S. 2016, Space Telescope WFC3
  Instrument Science Report

\bibitem[{Mack {et~al.}(2017)Mack, Lucas, Grogin, Koekemoer, \&
  Koekemoer}]{mack_2017}
Mack, J., Lucas, R., Grogin, N., Koekemoer, R. B. .~A., \& Koekemoer, A. 2017

\bibitem[{Mack {et~al.}(2021)Mack, Olszewksi, \& Pirzkal}]{mack_2021}
Mack, J., Olszewksi, H., \& Pirzkal, N. 2021, Space Telescope WFC3 Instrument
  Science Report

\bibitem[{{Madau} \& {Silk}(2005)}]{madau_2005}
{Madau}, P., \& {Silk}, J. 2005, \mnras, 359, L37,
  \dodoi{10.1111/j.1745-3933.2005.00031.x}

\bibitem[{{Matsumoto} {et~al.}(2018){Matsumoto}, {Tsumura}, {Matsuoka}, \&
  {Pyo}}]{matsumoto_2018}
{Matsumoto}, T., {Tsumura}, K., {Matsuoka}, Y., \& {Pyo}, J. 2018, \aj, 156,
  86, \dodoi{10.3847/1538-3881/aad0f0}

\bibitem[{Matsumoto {et~al.}(2005)Matsumoto, Matsuura, Murakami, Tanaka,
  Freund, Lim, Cohen, Kawada, \& Noda}]{matsumoto_2005}
Matsumoto, T., Matsuura, S., Murakami, H., {et~al.} 2005, The Astrophysical
  Journal, 626, 31–43, \dodoi{10.1086/429383}

\bibitem[{Matsumoto {et~al.}(2011)Matsumoto, Seo, Jeong, Lee, Matsuura,
  Matsuhara, Oyabu, Pyo, \& Wada}]{matsumoto_2011}
Matsumoto, T., Seo, H.~J., Jeong, W.-S., {et~al.} 2011, The Astrophysical
  Journal, 742, 124, \dodoi{10.1088/0004-637x/742/2/124}

\bibitem[{Matsuura {et~al.}(2017)Matsuura, Arai, Bock, Cooray, Korngut, Kim,
  Lee, Lee, Levenson, Matsumoto, Onishi, Shirahata, Tsumura, Wada, \&
  Zemcov}]{matsuura_2017}
Matsuura, S., Arai, T., Bock, J.~J., {et~al.} 2017, The Astrophysical Journal,
  839, 7, \dodoi{10.3847/1538-4357/aa6843}

\bibitem[{Maurer {et~al.}(2012)Maurer, Raue, Kneiske, Horns, Elsässer, \&
  Hauschildt}]{maurer_2012}
Maurer, A., Raue, M., Kneiske, T., {et~al.} 2012, The Astrophysical Journal,
  745, 166, \dodoi{10.1088/0004-637x/745/2/166}

\bibitem[{McKay \& Baggett(2017)}]{mckay_2017}
McKay, M., \& Baggett, S. 2017

\bibitem[{{Mihos}(2019)}]{mihos_2019}
{Mihos}, J.~C. 2019, arXiv e-prints, arXiv:1909.09456.
\newblock \doarXiv{1909.09456}

\bibitem[{Robitaille {et~al.}(2022)Robitaille, Sipőcz, Tollerud, Droettboom,
  Bradley, Bray, Lim, Deil, Craig, Price-Whelan, Barbary, Ginsburg, A., Horton,
  Morris, Spitler, Kerzendorf, Gupta, D'Avella, Burke, Crawford, Zabalza,
  danjampro, \& Günther}]{thomas_2022}
Robitaille, T., Sipőcz, B., Tollerud, E., {et~al.} 2022,
  AstroHuntsman/gunagala: v1.0.0, v1.0.0,  Zenodo,
  \dodoi{10.5281/zenodo.6796532}

\bibitem[{{Robotham} {et~al.}(2018){Robotham}, {Davies}, {Driver}, {Koushan},
  {Taranu}, {Casura}, \& {Liske}}]{robotham_2018}
{Robotham}, A.~S.~G., {Davies}, L.~J.~M., {Driver}, S.~P., {et~al.} 2018,
  \mnras, 476, 3137, \dodoi{10.1093/mnras/sty440}

\bibitem[{{Rowe} {et~al.}(2015){Rowe}, {Jarvis}, {Mandelbaum}, {Bernstein},
  {Bosch}, {Simet}, {Meyers}, {Kacprzak}, {Nakajima}, {Zuntz}, {Miyatake},
  {Dietrich}, {Armstrong}, {Melchior}, \& {Gill}}]{rowe15}
{Rowe}, B.~T.~P., {Jarvis}, M., {Mandelbaum}, R., {et~al.} 2015, Astronomy and
  Computing, 10, 121, \dodoi{10.1016/j.ascom.2015.02.002}

\bibitem[{{Rudick} {et~al.}(2011){Rudick}, {Mihos}, \& {McBride}}]{rudick_2011}
{Rudick}, C.~S., {Mihos}, J.~C., \& {McBride}, C.~K. 2011, \apj, 732, 48,
  \dodoi{10.1088/0004-637X/732/1/48}

\bibitem[{{Ryon}(2021)}]{acs_ihb}
{Ryon}, J.~E. 2021, in ACS Instrument Handbook for Cycle 29 v. 20.0, Vol.~20,
  20

\bibitem[{Sabbi \& Bellini(2013)}]{sabbi_2013}
Sabbi, E., \& Bellini, A. 2013, Space Telescope WFC3 Instrument Science Report

\bibitem[{Sano {et~al.}(2020)Sano, Matsuura, Yomo, \& Takahashi}]{sano_2020}
Sano, K., Matsuura, S., Yomo, K., \& Takahashi, A. 2020, The Astrophysical
  Journal, 901, 112, \dodoi{10.3847/1538-4357/abad3d}

\bibitem[{{Santos} {et~al.}(2002){Santos}, {Bromm}, \&
  {Kamionkowski}}]{santos_2002}
{Santos}, M.~R., {Bromm}, V., \& {Kamionkowski}, M. 2002, \mnras, 336, 1082,
  \dodoi{10.1046/j.1365-8711.2002.05895.x}

\bibitem[{Somerville {et~al.}(2012)Somerville, Gilmore, Primack, \&
  Domínguez}]{somerville_2012}
Somerville, R.~S., Gilmore, R.~C., Primack, J.~R., \& Domínguez, A. 2012,
  Monthly Notices of the Royal Astronomical Society, 423, 1992–2015,
  \dodoi{10.1111/j.1365-2966.2012.20490.x}

\bibitem[{{Stetson}(1987)}]{stetson87}
{Stetson}, P.~B. 1987, \pasp, 99, 191, \dodoi{10.1086/131977}

\bibitem[{{Symons} {et~al.}(2022){Symons}, {Zemcov}, {Cooray}, {Lisse}, \&
  {Poppe}}]{symons_2022}
{Symons}, T., {Zemcov}, M., {Cooray}, A., {Lisse}, C., \& {Poppe}, A.~R. 2022,
  arXiv e-prints, arXiv:2212.07449, \dodoi{10.48550/arXiv.2212.07449}

\bibitem[{{Unterborn} \& {Ryden}(2008)}]{unter08}
{Unterborn}, C.~T., \& {Ryden}, B.~S. 2008, \apj, 687, 976,
  \dodoi{10.1086/591898}

\bibitem[{van~de Hulst(1947)}]{hulst_1947}
van~de Hulst, H.~C. 1947, The Astrophysical Journal, 105, 471,
  \dodoi{10.1086/144921}

\bibitem[{{van der Wel} {et~al.}(2014){van der Wel}, {Franx}, {van Dokkum},
  {Skelton}, {Momcheva}, {Whitaker}, {Brammer}, {Bell}, {Rix}, {Wuyts},
  {Ferguson}, {Holden}, {Barro}, {Koekemoer}, {Chang}, {McGrath},
  {H{\"a}ussler}, {Dekel}, {Behroozi}, {Fumagalli}, {Leja}, {Lundgren},
  {Maseda}, {Nelson}, {Wake}, {Patel}, {Labb{\'e}}, {Faber}, {Grogin}, \&
  {Kocevski}}]{vanderwel14}
{van der Wel}, A., {Franx}, M., {van Dokkum}, P.~G., {et~al.} 2014, \apj, 788,
  28, \dodoi{10.1088/0004-637X/788/1/28}

\bibitem[{Virtanen {et~al.}(2020)Virtanen, Gommers, Oliphant, Haberland, Reddy,
  Cournapeau, Burovski, Peterson, Weckesser, Bright, {van der Walt}, Brett,
  Wilson, Millman, Mayorov, Nelson, Jones, Kern, Larson, Carey, Polat, Feng,
  Moore, {VanderPlas}, Laxalde, Perktold, Cimrman, Henriksen, Quintero, Harris,
  Archibald, Ribeiro, Pedregosa, {van Mulbregt}, \& {SciPy 1.0
  Contributors}}]{scipy_2020}
Virtanen, P., Gommers, R., Oliphant, T.~E., {et~al.} 2020, Nature Methods, 17,
  261, \dodoi{10.1038/s41592-019-0686-2}

\bibitem[{{Windhorst} {et~al.}(2011){Windhorst}, {Cohen}, {Hathi}, {McCarthy},
  {Ryan}, {Yan}, {Baldry}, {Driver}, {Frogel}, {Hill}, {Kelvin}, {Koekemoer},
  {Mechtley}, {O'Connell}, {Robotham}, {Rutkowski}, {Seibert}, {Straughn},
  {Tuffs}, {Balick}, {Bond}, {Bushouse}, {Calzetti}, {Crockett}, {Disney},
  {Dopita}, {Hall}, {Holtzman}, {Kaviraj}, {Kimble}, {MacKenty}, {Mutchler},
  {Paresce}, {Saha}, {Silk}, {Trauger}, {Walker}, {Whitmore}, \&
  {Young}}]{windhorst11}
{Windhorst}, R.~A., {Cohen}, S.~H., {Hathi}, N.~P., {et~al.} 2011, \apjs, 193,
  27, \dodoi{10.1088/0067-0049/193/2/27}

\bibitem[{{Windhorst} {et~al.}(2022){Windhorst}, {Carleton}, {O'Brien},
  {Cohen}, {Carter}, {Jansen}, {Tompkins}, {Arendt}, {Caddy}, {Grogin},
  {Koekemoer}, {MacKenty}, {Casertano}, {Davies}, {Driver}, {Dwek},
  {Kashlinsky}, {Kenyon}, {Miles}, {Pirzkal}, {Robotham}, {Ryan}, {Abate},
  {Andras-Letanovszky}, {Berkheimer}, {Chambers}, {Gelb}, {Goisman},
  {Henningsen}, {Huckabee}, {Kramer}, {Patel}, {Pawnikar}, {Pringle}, {Rogers},
  {Sherman}, {Swirbul}, \& {Webber}}]{windhorst_2022}
{Windhorst}, R.~A., {Carleton}, T., {O'Brien}, R., {et~al.} 2022, \aj, 164,
  141, \dodoi{10.3847/1538-3881/ac82af}

\bibitem[{{Windhorst} {et~al.}(2023){Windhorst}, {Cohen}, {Jansen}, {Summers},
  {Tompkins}, {Conselice}, {Driver}, {Yan}, {Coe}, {Frye}, {Grogin},
  {Koekemoer}, {Marshall}, {O'Brien}, {Pirzkal}, {Robotham}, {Ryan}, {Willmer},
  {Carleton}, {Diego}, {Keel}, {Porto}, {Redshaw}, {Scheller}, {Wilkins},
  {Willner}, {Zitrin}, {Adams}, {Austin}, {Arendt}, {Beacom}, {Bhatawdekar},
  {Bradley}, {Broadhurst}, {Cheng}, {Civano}, {Dai}, {Dole}, {D'Silva},
  {Duncan}, {Fazio}, {Ferrami}, {Ferreira}, {Finkelstein}, {Furtak}, {Gim},
  {Griffiths}, {Hammel}, {Harrington}, {Hathi}, {Holwerda}, {Honor}, {Huang},
  {Hyun}, {Im}, {Joshi}, {Kamieneski}, {Kelly}, {Larson}, {Li}, {Lim}, {Ma},
  {Maksym}, {Manzoni}, {Meena}, {Milam}, {Nonino}, {Pascale}, {Petric},
  {Pierel}, {del Carmen Polletta}, {R{\"o}ttgering}, {Rutkowski}, {Smail},
  {Straughn}, {Strolger}, {Swirbul}, {Trussler}, {Wang}, {Welch}, {B. Wyithe},
  {Yun}, {Zackrisson}, {Zhang}, \& {Zhao}}]{windhorst_2023}
{Windhorst}, R.~A., {Cohen}, S.~H., {Jansen}, R.~A., {et~al.} 2023, \aj, 165,
  13, \dodoi{10.3847/1538-3881/aca163}

\bibitem[{Wright(1998)}]{wright98}
Wright, E.~L. 1998, The Astrophysical Journal, 496, 1–8,
  \dodoi{10.1086/305345}

\bibitem[{Wright(2001)}]{wright_2001}
---. 2001, The Astrophysical Journal, 553, 538–544, \dodoi{10.1086/320942}

\bibitem[{Yue {et~al.}(2013)Yue, Ferrara, Salvaterra, Xu, \& Chen}]{yue_2013}
Yue, B., Ferrara, A., Salvaterra, R., Xu, Y., \& Chen, X. 2013, Monthly Notices
  of the Royal Astronomical Society, 433, 1556–1566,
  \dodoi{10.1093/mnras/stt826}

\end{thebibliography}
\bibliographystyle{aasjournal}
\appendix

\section{Simulated Images} \label{sec:simulations}

In order to develop and test reliable sky-SB measurement algorithms, we created simulated images where we know the true sky-SB and noise levels. We discuss here how these simulated images were created and the different kinds of simulated images that were produced.

\begin{figure}[h]
\epsscale{0.8}
\plotone{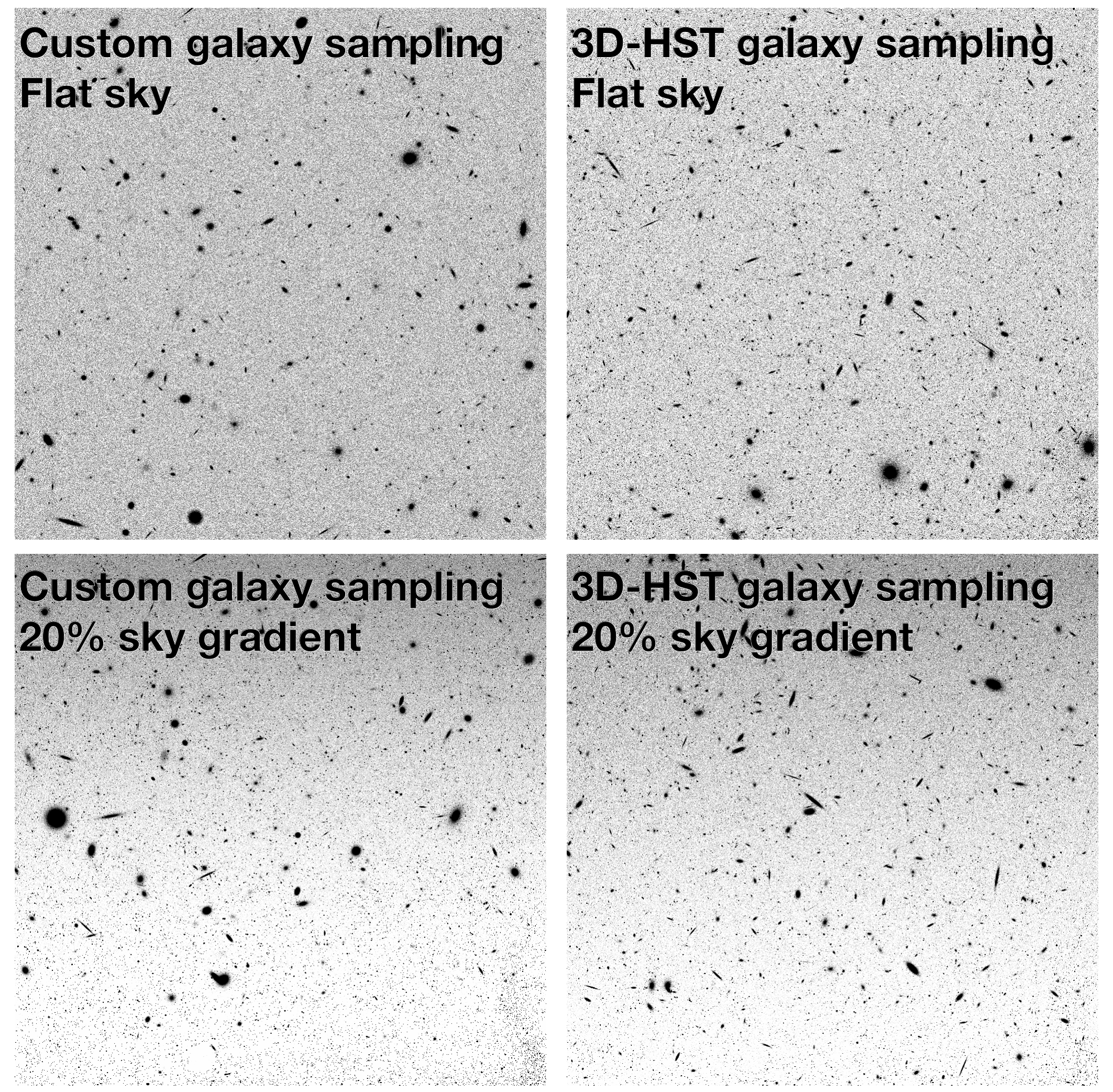}
\caption{Example simulated images. The custom galaxy sampling method (left column) employs a distribution of galaxy parameters based on \cite{windhorst11}. The COSMOS galaxy sampling method randomly selects galaxies from the COSMOS F125W database. The top images have no sky gradient added, while the bottom images have a sky gradient imposed on them.} \label{fig:simulatedim}
\end{figure}

We use GALSIM version 2.2.4 \citep{rowe15} to generate simulated images due to its ability to generate realistic galaxies and stars easily. These simulated images include stars, galaxies, cosmic rays, and sky gradients (see Figure \ref{fig:simulatedim}). All simulated images were produced to match WFC3/IR F125W flat-fielded images: $1014 \times 1014$ pixels, with a $0.13"$/pixel pixel scale. Therefore, star counts, galaxy counts, PSF sizes (necessary for GALSIM) and sky levels were also based on WFC3/IR F125W data. We produced a total of 344 images with a flat sky and 444 images with a sky gradient, with exposure times from 50s to 1302s, sky-SB levels ranging from 0.22 $e^-$ to 3.14 $e^-$, and sky gradients ranging from a 0\% change to a 20\% change edge-to-edge. We choose a wider range of sky-SB, noise, and sky gradient levels than is typical to ensure the robustness of our algorithms.

\subsection{Star and galaxy counts}

The star and galaxy counts for our simulated images are taken from \cite{windhorst11}. The star count slope for WFC3/IR data, shown in Equation \ref{eq:starcounts}, results in nearly 1 star per 1.0 mag bin within our chosen field of view. The number of stars in each simulated image is calculated as follows:

\begin{equation} \label{eq:starcounts}
    N_{\mathrm{stars},m_\mathrm{AB}} = 10^{0.04(m_\mathrm{AB}-18)}
\end{equation}

\noindent where $N_{\mathrm{stars},m_\mathrm{AB}}$ is the number of stars per integer AB magnitude ($m_\mathrm{AB}$) bin, where we assume an 18 mag bin contains exactly one star \new{\citep[as approximated from Figure 11 of][]{windhorst11}}.

Stars are restricted to $18 \leq \rm m_\mathrm{AB} \leq 26$ to avoid unusually bright stars and stars below the F125W detection limit. This resulted in a total of $\sim13$ stars generated in each simulated image. Every star was generated as a Gaussian with a full width at half maximum (FWHM) of $0.136"$. The position of each star in the simulated images was randomly selected with the condition that a star's center be within the $1014 \times 1014$ grid.

The galaxy count slope is steeper at around 0.26 dex/deg. The number of galaxies in each simulated image was calculated as follows:

\begin{equation} \label{eq:galcounts}
    N_{\mathrm{gal},m_\mathrm{AB}} = 10^{0.26(m_\mathrm{AB}-18)}
\end{equation}

\noindent where $N_{\mathrm{gal},m_\mathrm{AB}}$ is the number of galaxies per 0.5 AB magnitude bin.

Galaxies are restricted to $18 \leq \rm AB \leq 26.5$ to avoid unusually bright galaxies and galaxies below the F125W detection limit. This resulted in a total of $624$ galaxies generated in each simulated image. Every galaxy was generated using a single-component inclined S\'ersic profile (refer to \cite{rowe15} for profile details). Similarly to the simulated stars, the position of each galaxy in the simulated images was randomly selected with the condition that the galaxy's center be within the $1014 \times 1014$ grid.

The magnitude, effective radius, sersic index and axis ratio ($b/a$) were sampled using two methods: a custom distribution (described in Section \ref{sec:galsamp1}) and a random sampling from 3D-HST's COSMOS F125W Catalog (\cite{vanderwel14}; described in Section \ref{sec:galsamp2}).

\subsection{Galaxy Parameter Sampling Method 1: Custom Distribution} \label{sec:galsamp1}

The custom galaxy sampling method is motivated by \cite{windhorst11}. Images simulated with this method have the radius of each galaxy sampled from a distribution of the form:

\begin{equation} \label{eq:prob_galrad}
    p(R_e) = R_e\frac{e^{-R_e/0.2}}{0.2\times \Gamma(2)}
\end{equation}

\noindent where $p(R_e)$ is the probability density function for a galaxy with effective radius $R_e$, and $\Gamma$ is the Gamma function. This distribution follows closely with 3D-HST COSMOS F125W galaxy counts \citep{vanderwel14}. Due to GALSIM memory limitations, we only simulated galaxies with $R_e \leq 2.72''$. Galaxies with $R_e > 2.72''$ only account for $\ll 1\%$ of all galaxies in our field of view (follows from Equation \ref{eq:prob_galrad}).

S\'ersic indices for the custom sampled galaxies follow:

\begin{equation}
    p(n) = e^{0.38 n}
\end{equation}

\noindent where $p(n)$ is the probability density function for a galaxy with S\'ersic index $n$. Because the allowed range of S\'ersic indices for GALSIM is $0.3 \leq n \leq 6.2$, this is the range of S\'ersic index values present for the galaxies in the simulated images.

AB magnitudes for the custom sampled galaxies follow:

\begin{equation}
    p(m_\mathrm{AB}) = \frac{1}{\beta} \exp(\frac{m_\mathrm{AB}-26.5}{\beta}),
\end{equation}

\noindent where $\beta = \frac{1}{0.26 \times \ln(10)}$ and $p(m_\mathrm{AB})$ is the probability density function for a galaxy with AB magnitude $18<m_\mathrm{AB}<26.5$.

Lastly, the inclination of each galaxy produced from this method was randomly selected from the range $0$ to $\frac{\pi}{2}$ radians.

\subsection{Galaxy Parameter Sampling Method 2: 3D-HST COSMOS F125W Catalog} \label{sec:galsamp2}

Images simulated with this method used parameters directly sampled from the 3D-HST COSMOS F125W catalog \citep{vanderwel14}. In other words, galaxies were taken directly from the 3D-HST COSMOS F125W catalog and inserted into our simulated images. Inclinations were estimated using:

\begin{equation}
    \mathrm{cos}^2i=\frac{(b/a)^2 - \alpha^2}{1-\alpha^2}
\end{equation}

\noindent where $b/a$ is the axis ratio and $\alpha=0.22$ \citep{unter08}.

\subsection{Cosmic rays, noise and sky gradients} \label{sec:CRnoiseskygrad}

Cosmic rays (CRs) in the simulated images were generated by randomly selecting CRs from a WFC3/IR cosmic ray template and inserting them directly into the simulated images. The cosmic ray template was generated by identifying spikes in the individual reads of a random 1302 second HST image. This resulted in a rate of $14.6$ CRs per second over the course of the exposure. The number of CRs inserted into a simulated image is given by:

\begin{equation}
    N_{CRs} = R_{CR} \times t
\end{equation}

\noindent where $N_{CRs}$ is the number of cosmic rays in the simulated image, $R_{CR}$ is the CR rate of the cosmic ray template (14.6 CR's per second), and $t$ is the exposure time of the simulated image. 

Noise was generated for the simulated images using combination of Poisson noise (shot noise) and Gaussian read noise:

\begin{equation}
    \mathrm{RMS} = \frac{\sqrt{{S_{sky}} \times {t} + {RN}^2}}{t}
\end{equation}

\noindent where $S_{sky}$ is the sky background value, $t$ is the exposure time, and $RN$ is read noise. A read noise of 12 $e^-$ was used for all simulated images, and various different sky-SB and exposure time values were used. Poisson noise was added to the images first, followed by Gaussian read noise.

\new{Some simulated images were modeled with linear sky gradients. Real sky gradients may appear in images where an image is pushed too close to the Earth's limb, although these are often $<$10\% edge-to-edge. For this work, we include a sky gradient to also represent any type of light contamination that may appear in an image, such as Earth's limb, the extended halos of galaxies, faint stars, extended point spread functions, and optical ghosts. This allows us to test how well algorithms perform against \textit{any} source of stray light.} Sky gradients were generated according to:

\begin{equation}
    N_{row} = O_{row} + \frac{P}{100} \frac{R_{num}}{R_{tot}} O_{row}
\end{equation}

\noindent where $N_{row}$ are the gradient adjusted pixel values for a particular row of pixels in the image, $O_{row}$ are the non-gradient adjusted pixel values, $P$ is the percent change between the bottom and top row of the image, $R_{num}$ is the row number being adjusted, and $R_{tot}$ is the total number of rows in the image.

This method ensures that the true sky value in the gradient images is the lowest end of the gradient, aligning with our philosophy that the true sky value of a real HST image will have the least amount of light contamination and thus be the lowest sky value in an image.

\clearpage

\section{Choosing the most reliable algorithms}\label{sec:appendix_methods}

Nine independent sky-SB measurement algorithms were originally created for SKYSURF, with an end goal of using the most reliable and robust methods for SKYSURF. The chosen algorithms should be able to measure the true sky-SB accurately, even for cases with sky-SB contamination. As described in Appendix \ref{sec:simulations}, we create simulated images with sky gradients to simulate sky-SB contamination. \new{We present these algorithms here for users who might wish to implement their own sky-SB algorithms.}

\begin{figure*}[h]
\epsscale{1.18}
\plotone{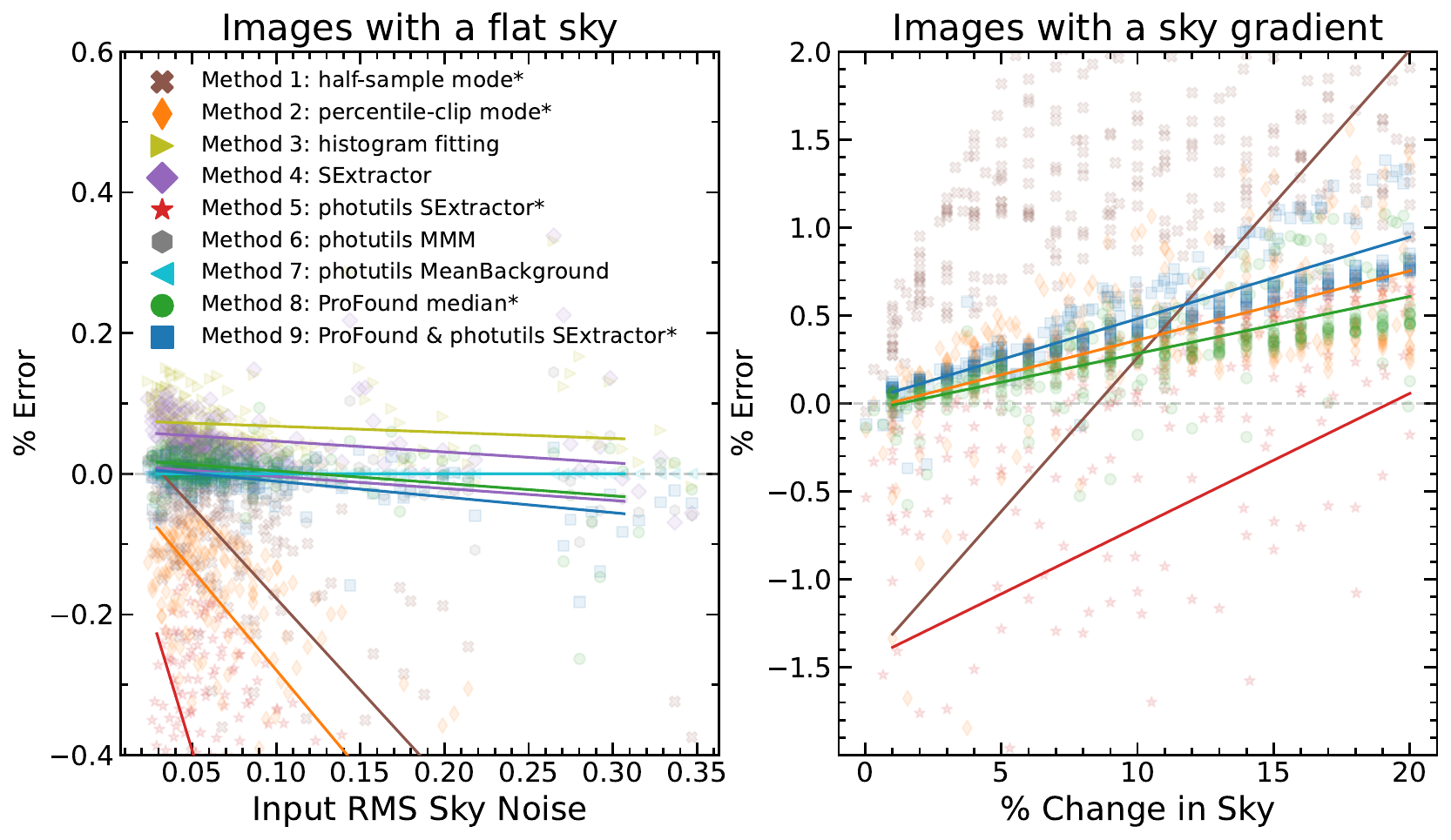}
\caption{Results of running various sky-SB estimation algorithms (Methods 1-9) on simulated HST WFC3/IR F125W images, where \% Error = (Measured Sky - Input Sky) / Input Sky $\times 100\%$. The solid colored lines represent the linear best fit for binned data where each bin contains 10-11 simulated images. Methods listed with an asterisk (*) are able to account for sky gradients, and are thus included in the right panel.
\textbf{Left:} Algorithm performance on simulated images with no sky gradient, plotted against the known, true sky-SB rms of the simulated image.
\textbf{Right:} Algorithm performance on simulated images with a sky gradient, plotted against the known sky gradient of the image.} \label{fig:methods}
\end{figure*}

The left-hand side of Figure \ref{fig:methods} shows that there are many methods that can retrieve the true input sky-SB to less than 0.1\% error. However, many of these methods (Methods 3, 4, 6, 7) assume a flat sky and can not account for sky gradients. We therefore developed several methods that account for sky gradients (Methods 1, 2, 5, 8 and 9; see right-hand plot in Figure \ref{fig:methods}). With SKYSURF's end goal of constraining a Diffuse Light signal, it is crucial that we are able to robustly ignore possible sources of sky-SB contamination. These could be sky gradients in the field of view due to Earth's limb, or the extended light profiles of galaxies. We choose the two algorithms that are able to best retrieve the known input sky-SB level from the simulated images with sky gradients: Method 2 (the Percentile-clip method) and Method 8 (the ProFound Median method).

The ProFound Median performs very well for flat images (less than 0.05\% error on average). The Percentile-clip method will underestimate the sky-SB in the case of a flat sky. The median sky-SB rms for WFC3/IR F125W images in our SKYSURF database is $\sim0.05$ electrons per second. Figure \ref{fig:methods} shows that the Percentile-clip method can still retrieve the known input sky-SB level to within 0.3\% for images with a flat sky and a sky-SB rms less than 0.05 electrons per second.

\new{These algorithms are extremely similar, as the Percentile-clip method uses a 5th-percentile to approximate the sky-SB, while the ProFound Median uses a median. In addition, the ProFound Median method uses ProFound SKY maps, while the Percentile-clip method does not. Besides these two differences, the algorithms are identical. We confirm that the use of ProFound SKY Maps typically results in a $\lesssim$0.5\% difference in sky-SB measurements when all other steps in the pipeline are the same, and decide to proceed with using ProFound SKY maps for \textit{both} methods. The new Percentile-clip method with ProFound SKY maps is named the ProFound-5th method. The main paper focuses on the ProFound Median method, while the ProFound-5th method is described in Appendix \ref{app:5th_method}.}

Here we describe each method in detail as it appears on Figure \ref{fig:methods}:

\begin{itemize}  

  \item[] \underline{Method 1: Half-sample mode} - 
  This method divides the image into $39\times39$ square pixel regions, then $3\sigma$ clips each sub-region. For each sub-region, it calculates the half-sample mode and rms, where the rms is determined to be the median absolute deviation multiplied by 1.48. This is able to estimate the mode of a sample by numerically finding the smallest interval that contains half of the points in a sample and iterating until obtaining an interval with only two points (e.g., \citeauthor{bickel05} \citeyear{bickel05}). The mode of the sample is approximated to be the average of the remaining two points. This method then identifies ``good'' cutout regions, assuming that the true (ZL+EBL) sky-SB of an image is closest to the LPS values in an image.
  
  \item[] \underline{Method 2: Percentile-clip} - 
  This method is described in Appendix \ref{app:5th_method}.
  
  \item[] \underline{Method 3: Histogram fitting} - 
  This method fits a histogram of sky-SB values from $-3\sigma$ to $+1\sigma$ using a clipped sigma as a first guess, and performs 2 iterations.
  
  \item[] \underline{Method 4: SourceExtractor} - 
  This method uses \verb|SourceExtractor| \citep{bertin96} to calculate the sky-SB, where we assume that each object has a different surrounding sky.
  
  \item[] \underline{Method 5: photutils SourceExtractor} - This method masks all sources in an image then splits each image into $26\times26$ square pixel regions with a 30 pixel border surrounding each image that is subsequently ignored. It estimates the sky-SB value of each cutout using the \verb|photutils SourceExtractor| algorithm \citep{bradley20}. The sky-SB rms is estimated using the \verb|photutils| median absolute deviation algorithm. It rejects sub-regions with a measured sky-SB greater than the lowest sky-SB $+$ the average sky-SB rms of all sub-regions. It also rejects sub-regions with a measured sky-SB rms greater than twice the average rms of all sub-regions. This method then estimates the sky-SB gradient of the image using the brightest 7\% of cutout regions and the dimmest 7\% of cutout regions. Using this calculated gradient, the algorithm determines a threshold (N) for which to include images in the final calculation, where N ranges from 4\% to 35\% for large to small gradients, respectively. The sky-SB of the image is the lowest N\% of good cutout regions.
  
  \item[] \underline{Method 6: photutils MMM} - 
  This method masks all the sources in the image, then calculates the sky-SB using a \verb|photutils| method based on the \verb|DAOPHOT MMM| algorithm \citep{bradley20, stetson87}.
  
  \item[] \underline{Method 7: photutils Mean} - 
  Method 7 masks the sources in an image, $1.3\sigma$ clips the entire image, then calculates the sky-SB using \verb|photutils MeanBackground|.
  
  \item[] \underline{Method 8: ProFound median} - This method is described in Section \ref{sec:methods}.
  
  \item[] \underline{Method 9: ProFound \& photutils Source-Extractor} - This method utilizes Method 5 (photutils SourceExtractor) on ProFound SKY maps.
  
\end{itemize}

\clearpage

\section{The ProFound-5th Algorithm} \label{app:5th_method}

As stated in \citetalias{windhorst_2022}, we measure the sky-SB for all images in our database using our two best algorithms. In this section, we describe the second algorithm that is not described in the main paper: the ProFound-5th (Pro-5th) method. It follows the same methodology as the Pro-med method, but uses a 5th-percentile of unflagged sub-regions rather than a median. In other words, the final sky-SB level of a chip, $\skyim$, is the 5th-percentile of the remaining $\skysub$ values. The final sky-SB rms of a chip, $\skyrmsim$, is the mean of all the $\skysubrms$ values.

In contrast to the Pro-med method, the Pro-5th method recovers the darkest measurable sky-SB for every HST image. Many SKYSURF images contain light from the extended profiles of galaxies \citep{ashcraft_2018a}, the extended point-spread functions (PSFs) of stars \citep{borlaff_2019}, thermal foregrounds \citepalias{carleton_2022}, and the Earth's atmosphere \citep{caddy_2022}. While explicitly modeling and subtracting these features of the measured sky-SB is out of the scope of SKYSURF, we reduce the probability of unaccounted-for sources of flux that may contaminate our sky-SB measurements by choosing to isolate the darkest area in an image.

\begin{figure*}[t]
    \centering
    \includegraphics[scale = 0.25]{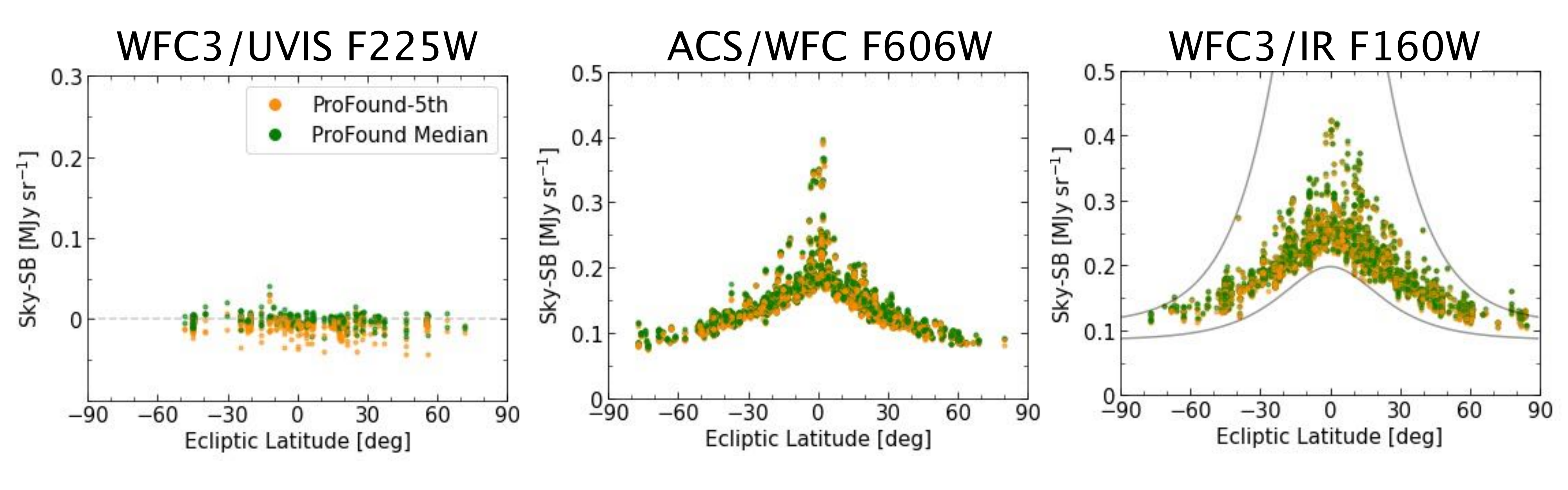}
    \caption{SKYSURF sky-SB measurements versus Ecliptic Latitude for three example filters. We compare the Pro-med method (green) to the Pro-5th method (orange). As an example of the sky-SB measurements we expect, we include the WFC3/IR F160W $sech$ curve from Figure 2 of \citetalias{carleton_2022}, which is derived to match Kelsall COBE/DIRBE model predictions. Our measurements fall conservatively within these limits.}
    \label{fig:sky_vs_ecllat_both}
\end{figure*}

As shown in Figure \ref{fig:sky_vs_ecllat_both}, the Pro-5th measurements fall below zero in UV wavelengths. \new{The lack of objects (because hot stars are rare) in this wavelength range means that Pro-5th method more severely underestimates sky than in longer wavelengths.} In addition, the measurement noise in UV wavelengths is larger than at other wavelengths, as shown in Section \ref{sec:skyrms}. This effect will also cause the Pro-5th method to be more biased with respect to the true value. Figure \ref{fig:methods} shows that a 5th-percentile method (method \#2) underestimates the true sky-SB by a larger margin as the sky-SB rms increases.

Following the same methods of Section \ref{sec:diffuse_light_limits}, we calculate DL limits for the Per-clip algorithm. We estimate DL limits of 0.008 MJy/sr for F125W, 0.015 MJy/sr for F140W, and 0.012 MJy/sr for F160W. This method still provides DL limits that are in excellent agreement (to within error) to the Pro-med method.

\subsection{Comparison of the two methods} \label{sec:comparison_of_methods}

\begin{figure*}[h]
    \centering
    \includegraphics[scale = 0.42]{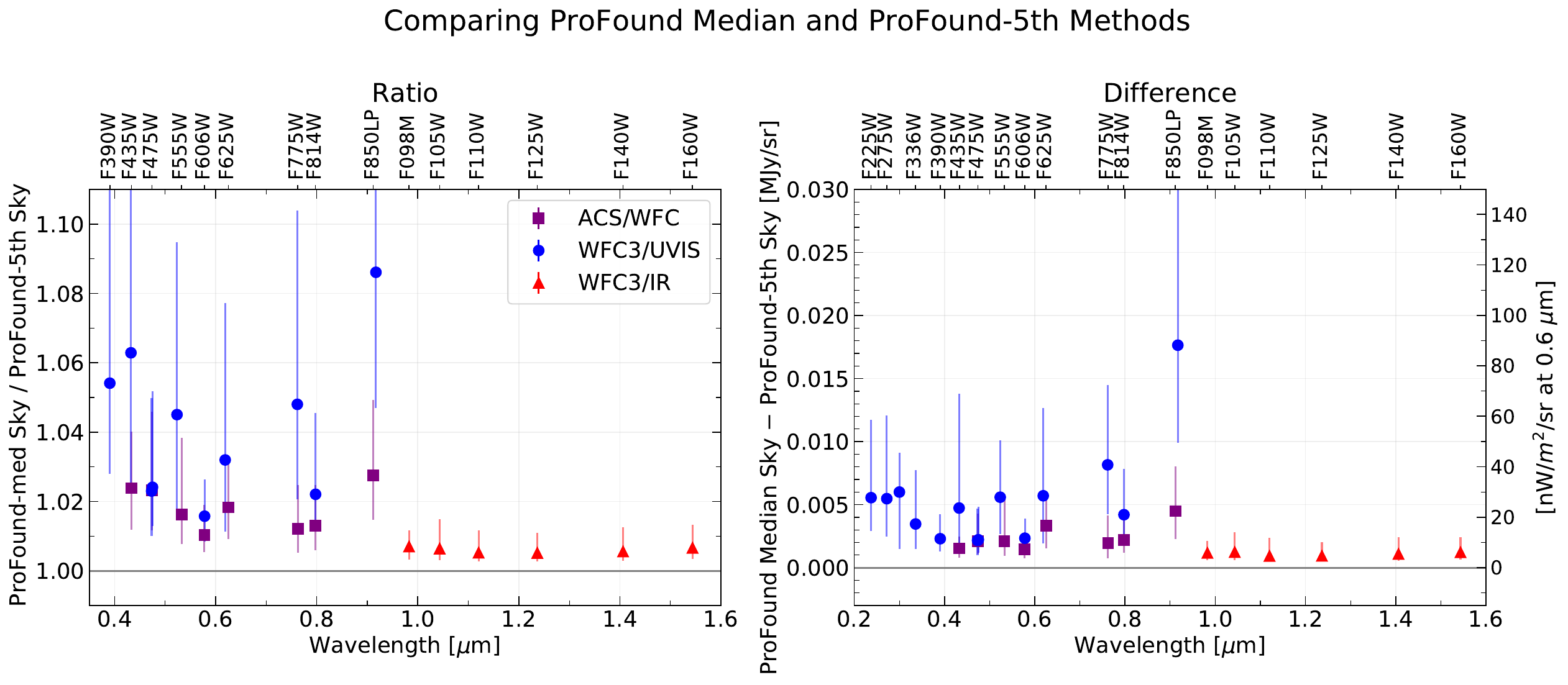}
    \caption{Comparison of the Pro-med method and the Pro-5th algorithms. The error bars show the 16th- and 84th-percentiles of the y-axis distributions. \textbf{Left:} Median ratio of the Pro-med sky-SB divided by the Pro-5th sky-SB. The bluest filters are excluded because the sky-SB is nearly zero, such that the ratios become extreme values. \textbf{Right:} Median difference in Pro-Med and Pro-5th method. The left y-axis shows units of \MJysr and the right axis shows \nWmsr{} at 0.6 \micron.}
    \label{fig:comparing_percentileclip_to_profound}
\end{figure*}

We compare the sky-SB levels of the Pro-med and Pro-5th algorithms for each HST filter. As shown in Figure \ref{fig:comparing_percentileclip_to_profound}, the Pro-med method gives sky-SB values that are on average 2\% higher for ACS/WFC, 4\% higher for WFC3/UVIS, and 0.6\% higher for WFC3/IR than the Pro-5th method. \new{This is an expected difference, as the Pro-5th method will always result in a lower sky-SB because it probes the darkest part of an image.}

The trend seen as a function of wavelength in Figure \ref{fig:comparing_percentileclip_to_profound} is due to trends of the average sky-SB rms (Figure \ref{fig:rms_vs_wave}). \new{This can be inferred from the fact that a lower sky-SB rms results in a smaller variation in the sky pixels. In other words, a lower average sky-SB rms results in a smaller difference between methods because the difference between the median and the 5th-percentile of a gaussian distribution is smaller when the rms is lower. Although these trends are driven by Poisson statistics, other factors that could affect the sky-SB rms include increased contamination from contaminants or sky gradients.}

\clearpage

\section{Sources of Uncertainty} \label{sec:error_appendix}

In this section, we expand on the uncertainties present in Section \ref{sec:error}.

\subsection{WFC3/UVIS Uncertainties} \label{subsec:error_wfc3uvis}

Flat-field errors are $\leq$1\% \citep{mack_2016}. However, errors can be larger in the corner of the UVIS1 chip where the point-spread function focus is highly variable due to the telescope breathing effects \citep{sabbi_2013}, and this impacts the flat field correction. Also, small offsets between the two different WFC3/UVIS detectors are present in some filters, with a maximum difference between one corner of a detector to the other of $\sim$3\% \citep{mack_2016} for a few ultraviolet filters. These variations in flat-field could potentially bias our results. In Appendix \ref{app:trends_in_darkest_subregions_on_the_ccd}, we independently explore maximum possible uncertainties in the flat-fields by taking advantage of the large SKYSURF database. We do this by comparing the systematically darkest and brightest sub-regions, with a typical offset of 2--4\%, which agrees with \citet{mack_2016}. This is a maximum difference between the darkest and brightest sub-regions, but our algorithms automatically ignore the very darkest and brightest sub-regions. We therefore adopt the \citet{mack_2016} flat-field uncertainty of 1\%.

We use the new photometric zeropoint calibrations explained in \cite{calamida_2022}, where they account for variations in WFC3/UVIS zeropoints over time. We adopt the photometric errors listed in Table 8 of \cite{calamida_2022}, which on average represent a $<$0.2\% 1$\sigma$ dispersion (Table 8 of \citeauthor{calamida_2022} \citeyear{calamida_2022}).

A bias offset is added to HST detectors to avoid presenting a negative voltage to the analog-to-digital converter. This offset is always subtracted during post-processing, and uncertainty in the bias level introduces error to the sky-SB.
Figure 5 of \cite{mckay_2017} shows a scatter in individual bias levels of $\sim0.2$ electrons. Figure 17 from \cite{bourque_2016} shows that the scatter in determining the dark current is $\sim$2 e-/hr or $\sim$0.0006 electrons per second.

Different versions of the standard WFC3 calibration pipeline correct for CTE trails differently. We quantify the effects of different pipeline versions in Appendix \ref{app:comparing_calwf3_versions}. We find the measured sky-SB between different versions of the pipeline to be $\sim0.007$\% for wavelengths longer than 0.4 microns. We therefore do not include CTE effects in our error budget. As discussed in \citetalias{windhorst_2022}, we adopt a post-flash subtraction error of 1\%, corresponding to 0.16 electrons for a F606W image with an exposure time of 500 seconds.

\subsection{WFC3/IR Uncertainties}


\cite{mack_2021} present residuals in the sky flats of 0.5--2\%. We adopt a conservative WFC3/IR flat field uncertainty of 2\%. As described in \citetalias{windhorst_2022}, WFC3/IR photometric zeropoints have roughly remained constant to within 1.5\% (rms) since 2009. Therefore, we adopt a zeropoint uncertainty of 1.5\% for WFC3/IR.

As described in \citetalias{wfc3_ihb}, the WFC3/IR detector responds non-linearly to incident photons. The WFC3 calibration pipeline corrects for this with a $\sim0.5$\% uncertainty. We therefore adopt a $\sim0.5$\% uncertainty in the non-linearity of WFC3/IR. WFC3/IR detector artifacts, most notably the IR blobs, are ignored by masking corresponding pixels flagged in the DQ array.

As described in \citetalias{windhorst_2022}, we adopt a dark/ bias uncertainty of 1\% for WFC3/IR, corresponding to 0.005 electrons per second for a F125W image. We define the thermal dark signal to be thermal noise from the telescope assembly and instruments \citepalias[see][]{carleton_2022}. It is strongly dependent on wavelength, where it is negligible below 1 \micron{} and significant above 1.4 \micron{}. As shown in \citetalias{carleton_2022}, the maximum error we expect is 2.7\% for F160W, with lower uncertainties for F125W and F140W. To be conservative, we adopt 0.01 electrons per second uncertainty in the thermal dark signal for all WFC3/IR sky-SB measurements in this report. Carleton et al. (in prep) will provide better constraints on the thermal dark signal.

During manual inspection of images, we noticed clear amplifier offsets. These effects are known to be due to differences in the noise and gain between amplifiers. In Appendix \ref{app:amplifier_differences}, we explore the effect this has on sky-SB estimates. We find median differences in pixel column values close to the amplifier boundaries to be $<0.2$\% for all WFC3/IR filters. We therefore do not include amplifier differences in our WFC3/IR error budget.

\subsection{ACS/WFC Uncertainties}

\cite{cohen_2020} find that the newest ACS/WFC flat-fields result in a photometric scatter of point sources that range from 0.5\% to 3\%. They claim this could be contributions from various reference files and CTE losses that are underestimated. In Appendix \ref{app:trends_in_darkest_subregions_on_the_ccd}, and mentioned in Section \ref{subsec:error_wfc3uvis}, we independently explore uncertainties in the flat-fields by taking advantage of the large SKYSURF database. On average, our results agree with \cite{cohen_2020}. Following \cite{windhorst_2022}, we adopt the conservative uncertainty in the ACS/WFC flat field to be 2.2\%. As described in \citetalias{windhorst_2022}, we adopt a zeropoint uncertainty for ACS/WFC to be 1\% (Figure 2 \citeauthor{bohlin_2020} \citeyear{bohlin_2020}).

ACS/WFC exhibits bias offsets that vary from amplifier to amplifier, as described in \citet{acs_dhb}. The accuracy of the bias level subtraction is limited by random variations of about 0.3 DN (0.6 electrons). As described in \citetalias{windhorst_2022}, the ACS/WFC exhibits dark current uncertainty of ~0.001 electrons per pixel per second. Figure 3 from \cite{anand_2022} shows a scatter in the ability to determine the ACS/WFC dark current to be $\sim$0.001 electrons per second (0.5 electrons for a 500 second exposure). Finally, we adopt a postflash uncertainty for ACS/WFC to be 0.37 electrons \citepalias{windhorst_2022}.

\subsection{Crosstalk} \label{sec:crosstalk}

The CCD's on ACS are known to suffer from crosstalk \citep[][]{giavalisco_2004}, where artificial ``ghosts'' from bright objects appear in mirror-symmetric positions in other ACS quadrants. These ghosts will appear as depressions relative to the background, with strengths of only a few electrons per pixel. However, the sky-SB itself can be on the order of a few electrons per pixel, so it is necessary we take crosstalk into consideration. Therefore, we correct for crosstalk to see how it affects our sky-SB measurements. For this simple test, we focus on ACS F775W images, where we analyze each ACS chip independently. Our default algorithm masks sky sub-regions that likely contain discrete objects. To correct for crosstalk, we reflect all masked regions to the opposite quadrant to mask the ghost corresponding to any bright objects. After reflecting masked regions, we recalculate the sky-SB. We find that crosstalk affects the sky-SB by only 0.008\% for 68\% of measurements. 95\% of the measurements have a 0.08\% difference. We conclude that our algorithm is robust enough that crosstalk does not significantly impact sky-SB levels.

\subsection{Amplifier Differences} \label{app:amplifier_differences}

The readout amplifiers can introduce additional errors to our analysis. Differences in readout noise and gain between amplifiers can cause artificial variations in the sky-SB level. Differences in sky-SB will appear as additive differences in the background for the part of the detector that is read out to a corresponding amplifier. Correcting for this without affecting the true sky-SB level is very difficult. It requires identification of differences in background level exactly at the amplifier boundaries without taking light from objects into account. This would mean only using pixels close to the amplifier boundaries, many of which are likely contaminated by discrete objects. We consider this task to be past the scope of this project. However, we test for systematic differences across our database by measuring how the mean pixel value of a pixel column varies across amplifier boundaries. The median difference between the 20 pixel columns to the left of the amplifier boundary and the 20 pixel columns to the right of the amplifier boundary is always $<0.2$\% for WFC3/IR images in our database. We therefore do not include amplifier differences into our uncertainty estimations.

\subsection{Trends in darkest sub-regions on the CCD} \label{app:trends_in_darkest_subregions_on_the_ccd}

If there are regions on a detector that have systematically lower sky-SB values due to flat-field error, bias error, or geometrical distortion, this could potentially bias our results.  \new{We run our algorithm (described in Section \ref{sec:methods}) using \flt{}/\flc{} files as input instead of ProFound SKY maps, as ProFound can smooth over the CCD structures we want to study (i.e. flat-field and distortion).} We test for systemically darker sub-regions by creating two-dimensional histograms of the darkest 5\% of sub-regions from our sky estimation algorithms. We refer to these as the ``darkest sub-regions'' for this discussion.

\begin{figure}[h]
    \centering
    \includegraphics[scale = 0.5]{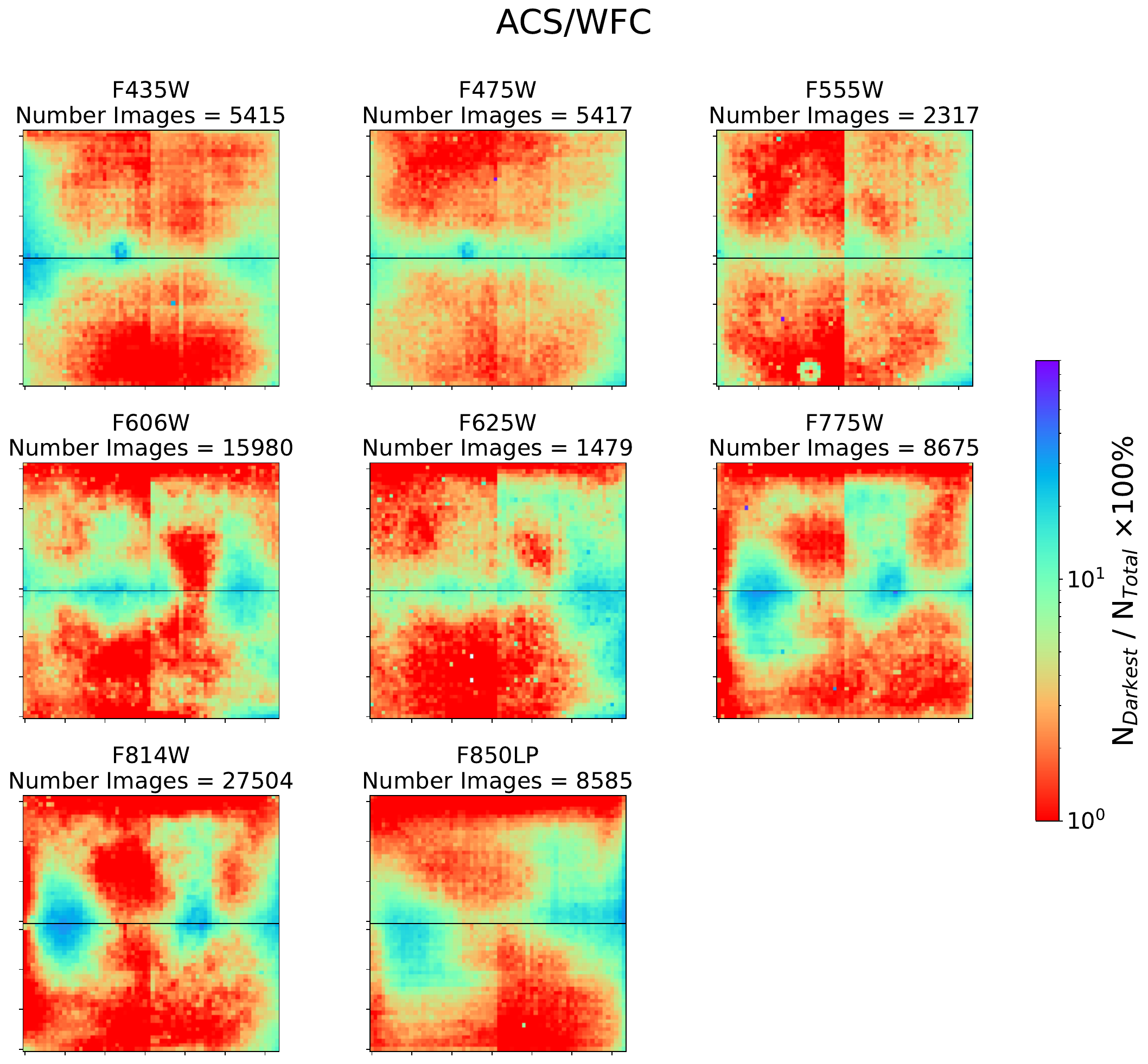}
    \caption{Two-dimensional histograms of the darkest sub-regions for all the images in every filter in ACS/WFC. The colorbar indicates the percent of images ($N_{\text{Darkest}} / N_{\text{Total}} \times 100\%$ where $N_{\text{Darkest}}$ is the number of darkest sub-regions and $N_{\text{Total}}$ is the total number of images) in this filter that contain a darkest sub-region in the corresponding location. In other words, regions that are more blue/ purple have systematically lower sky-SB levels across the entire filter. These histograms resemble the ACS/WFC flat-fields, indicating a residual flat-field exists in a majority of images in our database.}
    \label{fig:acswfc_stacked_green}
\end{figure}

\begin{figure}[h]
    \centering
    \includegraphics[scale = 0.4]{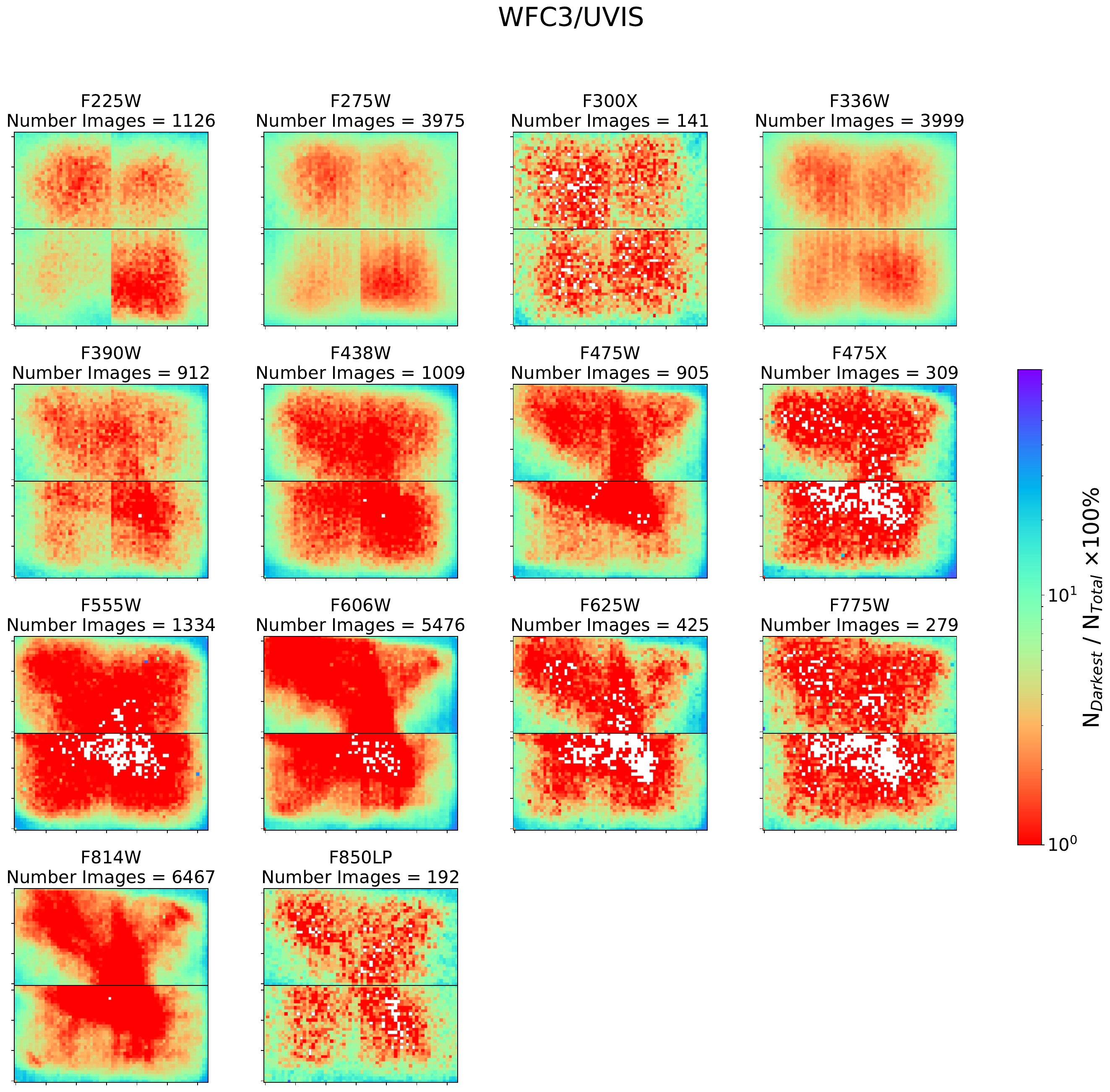}
    \caption{Two-dimensional histograms of the darkest sub-regions for all the images in every filter in WFC3/UVIS. The colorbar indicates the percent of images ($N_{\text{Darkest}} / N_{\text{Total}} \times 100\%$ where $N_{\text{Darkest}}$ is the number of darkest sub-regions and $N_{\text{Total}}$ is the total number of images) in this filter that contain a darkest sub-region in the corresponding location. In other words, regions that are more blue/ purple have systematically lower sky-SB levels across the entire filter. White indicates regions that are always ignored due to detector artifacts or regions that always contain a bright object. These histograms resemble the WFC3/UVIS flat-fields, indicating a residual flat-field exists in a majority of images in our database.}
    \label{fig:wfc3uvis_stacked_green}
\end{figure}

\begin{figure}[h]
    \centering
    \includegraphics[scale = 0.38]{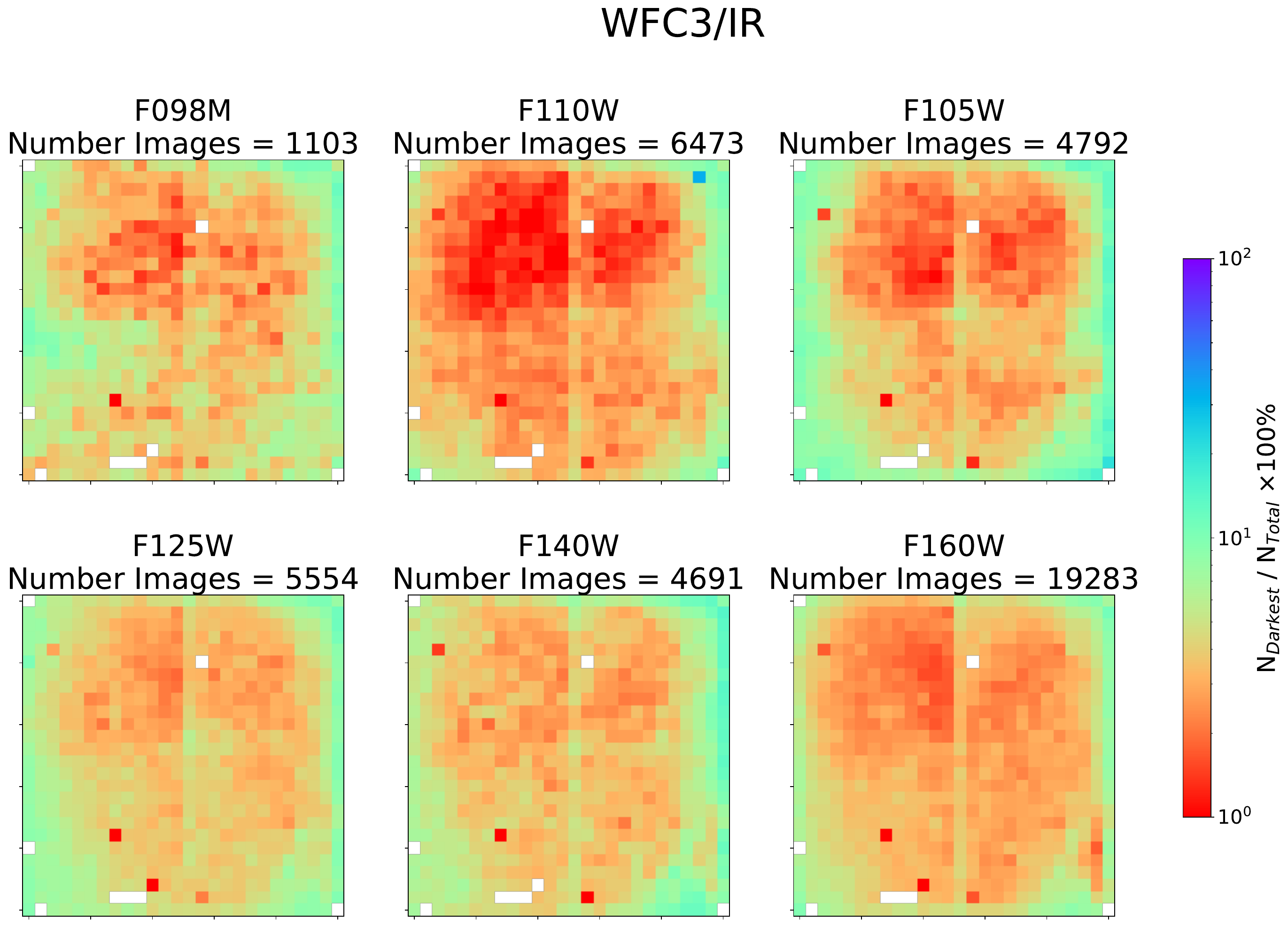}
    \caption{Two-dimensional histograms of the darkest sub-regions for all the images in every filter in WFC3/IR. The colorbar indicates the percent of images ($N_{\text{Darkest}} / N_{\text{Total}} \times 100\%$ where $N_{\text{Darkest}}$ is the number of darkest sub-regions and $N_{\text{Total}}$ is the total number of images) in this filter that contain a darkest sub-region in the corresponding location. In other words, regions that are more blue/ purple have systematically lower sky-SB levels across the entire filter. White indicates regions that are always ignored due to detector artifacts. Known artifacts that are always masked are the ``death star'' (bottom middle) and ``wagon wheel'' (bottom right corner).}
    \label{fig:wfc3ir_stacked_green}
\end{figure}

The resulting histograms are shown in Figures \ref{fig:acswfc_stacked_green}-\ref{fig:wfc3ir_stacked_green}. To best probe CCD structure, we use \textit{all} images in our SKYSURF database. Each two-dimensional histogram includes $N\simeq 1000-27,000$ images, so any subtle effects on the sky-SB from a particular detector location can be sampled this way. Bluer boxes indicate regions where most SKYSURF images contain a darkest sub-region in the corresponding location on the detector. In other words, bluer regions are systematically darker. Redder boxes indicate regions where most SKYSURF images do not contain a darkest subregion in the corresponding location on the chip. For ACS/WFC and WFC3/UVIS, there is a clear structure in the histograms that resemble the flat-fields \citep{mack_2017,wfc3_ihb}. The structure for WFC3/IR is less obvious, where the darkest sub-regions instead tend along the WFC3/IR amplifier boundaries.

\new{The residual flat-field seen in Figures \ref{fig:acswfc_stacked_green}-\ref{fig:wfc3ir_stacked_green} could potentially affect sky-SB measurements. Because these figures resemble the flat fields, we use them to quantify uncertainties in flat field estimates.} We use the histograms in Figures \ref{fig:acswfc_stacked_green}-\ref{fig:wfc3ir_stacked_green} to find the systematically darkest (darkest 5\% of sub-regions) and brightest regions (brightest 5\% of sub-regions). For every SKYSURF image in a filter, we take the mean sky-SB of the previously identified darkest regions and the mean sky-SB of the brightest regions. Note that these regions are the \textit{same} for every image in a filter because they are determined using Figures \ref{fig:acswfc_stacked_green}-\ref{fig:wfc3ir_stacked_green}.

\begin{figure}[h]
    \centering
    \includegraphics[scale = 0.42]{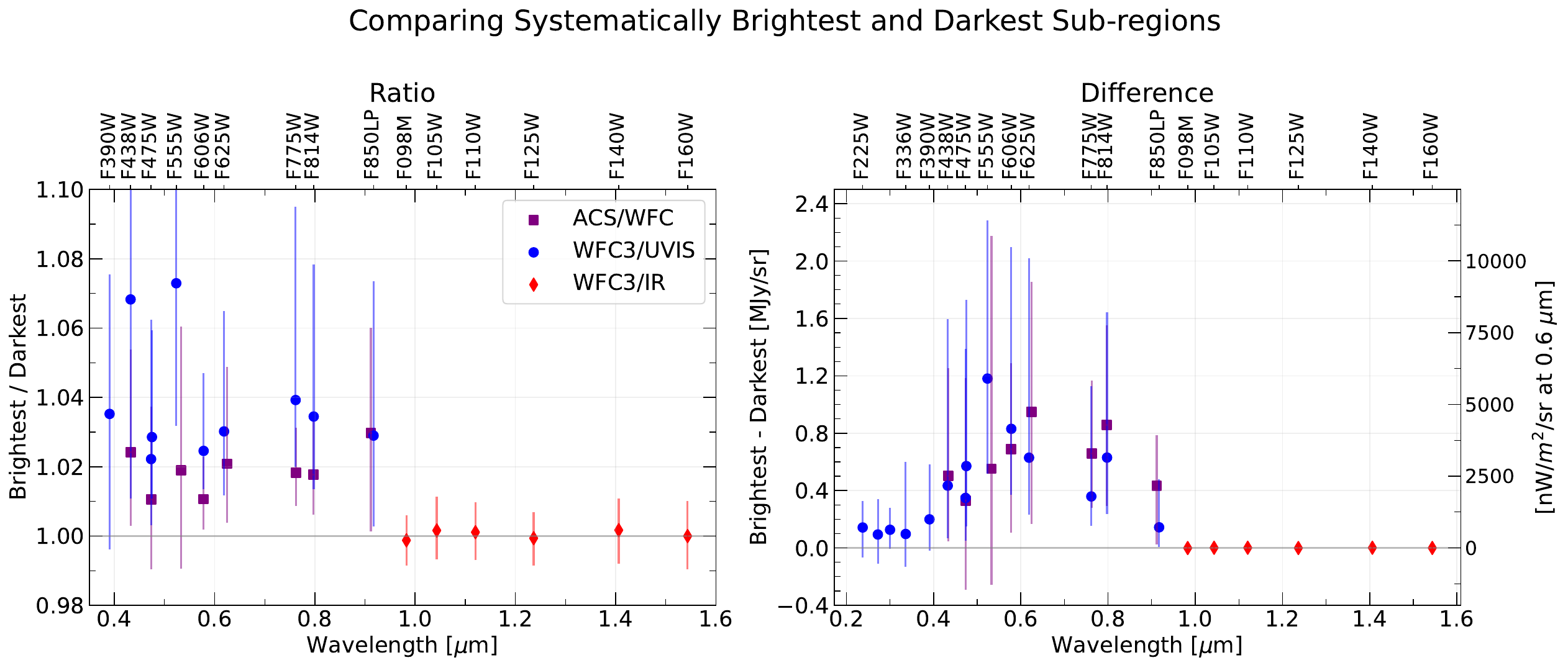}
    \caption{Comparison of the systematically brightest and darkest sub-regions. The error bars for both plots show the 16th- and 84th-percentiles of the y-axis distribution. \textbf{Left:} Median ratio of the brightest sub-regions over the darkest sub-regions. The bluest filters are excluded because the sky-SB is nearly zero. \textbf{Right:} Median difference of the brightest sub-regions over the darkest sub-regions. The left y-axis shows units of \MJysr{} and the right axis shows \nWmsr{} at 0.6 \micron.}
    \label{fig:brightest_darkest_subregions}
\end{figure}

We can compare the mean of the darkest and brightest regions to estimate the maximum level of uncertainty that the structure seen in Figures \ref{fig:acswfc_stacked_green}-\ref{fig:wfc3ir_stacked_green} add to surface brightness measurements. Figure \ref{fig:brightest_darkest_subregions} shows the ratio of the brightest and dimmest sub-regions. We find that WFC3/IR does not show significant sky-SB differences between the darkest and brightest regions. However, WFC3/UVIS and ACS/WFC tend to show $\sim2-5$\% differences.

This is a unique test of flat-field uncertainty due to SKYSURF’s massive database and corresponding sky-SB measurements. Although not necessarily relevant for this work, this test gives a maximum possible flat-field uncertainty on aperture photometry. We still adopt \cite{mack_2016} as our flat field uncertainty for SKYSURF sky-SB measurements.

\clearpage

\subsection{Testing how different CTE corrections affect the WFC3/UVIS sky-SB} \label{app:comparing_calwf3_versions}

As described in Appendix B.2 of \citetalias{windhorst_2022}, we redownloaded WFC3/UVIS images calibrated with the newest \texttt{calwf3} version at the time: \texttt{calwf3} v3.6.0. This version of the standard calibration pipeline presents updates to the CTE corrections. Since CTE corrections adjust pixels containing a sky-SB signal, we ensure these updated corrections do not significantly affect sky-SB measurements. Results are shown in Table \ref{tab:comparing_calwf3_versions}. The sky-SB rms improves for all filters, with sky-SB rms values typically being 2--5\% lower with the v3.6.0 corrections. We find an average median offset in sky-SB between both pipelines of $\sim0.007$\% for filters longward of 0.4 \micron{} and an average median offset between both pipelines of $\sim1.9$\% for filters shortward 0.4 \micron{}.

\begin{table}[h]
\centering
\begin{tabular}{|ccc|}
\hline
Filter &  Median Sky Ratio &  Median Sky rms Ratio \\
\hline
 F225W &          0.979750 &              0.978251 \\
 F275W &          0.877970 &              0.969360 \\
 F300X &          1.037716 &              0.952036 \\
 F336W &          1.014963 &              0.979252 \\
 F390W &          0.997349 &              0.975110 \\
 F438W &          1.002838 &              0.975508 \\
 F475X &          1.008633 &              0.964548 \\
 F475W &          1.000642 &              0.969892 \\
 F555W &          0.999497 &              0.970605 \\
 F606W &          1.001499 &              0.958634 \\
 F625W &          0.999005 &              0.962439 \\
 F775W &          0.987096 &              0.973376 \\
F850LP &          0.999113 &              0.972571 \\
 F814W &          1.001019 &              0.963322 \\
\hline
\end{tabular}
\caption{Median ratio of the calwf3 v3.5.0 / calwf3 v3.6.0 sky-SB values and sky-SB rms values. Only reliable sky-SB measurements are used in this comparison.}
\label{tab:comparing_calwf3_versions}
\end{table}

\section{Converting to Flux Units} \label{app:converting_to_flux_units}

Given the confusion that sometimes arises on this topic, in this section we explain our methods to convert our sky-SB measurements to units of spectral flux density. The conversions for each camera are highlighted in Equations \ref{eq:wfc3ir_flux_calculation}-\ref{eq:acswfc_flux_calculation}. We adopt the pixel areas described in the WFC3 Instrument Handbook \citep{wfc3_ihb} and the ACS Instrument handbook \citep{acs_ihb}: $0.135\times0.121$ arcsec$^2$ for WFC3/IR, $0.0395\times0.0395$ arcsec$^2$ for WFC3/UVIS and $0.050\times0.050$ arcsec$^2$ for ACS/WFC.



\begin{equation} \label{eq:wfc3ir_flux_calculation}
    I_{\lambda, \text{WFC3/IR}} = 
    \frac{\skyim\times\texttt{PHOTFNU}}
    {A}
\end{equation}

\begin{equation} \label{eq:wfc3uvis_flux_calculation}
    I_{\lambda, \text{WFC3/UVIS}} = 
    \frac{\skyim \times \texttt{PHTFLAM}\text{(converted)}}
    {A}
\end{equation}

\begin{equation} \label{eq:acswfc_flux_calculation}
    I_{\lambda, \text{ACS/WFC}} = 
    \frac{\skyim \times \texttt{PHTFLAM}\text{(converted)}}
    {A}
\end{equation}

For WFC3/IR in Equation \ref{eq:wfc3ir_flux_calculation}, $\skyim$ is the measured sky-SB in units of electrons per second ($e^-$/s), \texttt{PHOTFNU} is the inverse sensitivity taken from the image header in units of Jy/($e^-$/s), and $A$ is the average pixel area in units of steradians. When using SKYSURF sky-SB measurements for SKYSURF, a thermal dark signal must be subtracted \citep{carleton_2022}. Please refer to Carleton et al. (in prep) for updated estimates on thermal dark levels.

\new{For all images in our database, a flux correction (indicated by keyword \texttt{FLUXCORR}) is performed to bring the sensitivity of UVIS2 (one of the WFC3/UVIS chips) to the same sensitivity of UVIS1. However, the \texttt{PHOTFNU} keyword does not update with this change. Instead, the the chip-dependent inverse sensitivity, \texttt{PHOTFLAM}, is updated, so we use this keyword instead. \texttt{PHTFLAM} is used for both WFC3/UVIS and ACS/WFC,} where we convert it from units of erg cm$^{-2}$ \AA$^{-1}$ s$^{-1}$ to units of Jy (\texttt{PHTFLAM}(converted)) using the chip-dependent pivot wavelength listed in the header (\texttt{PHTPLAM}). This conversion is done using \texttt{astropy.units} \citep{astropy:2013, astropy:2018}.

\section{Ensuring A Reliable Diffuse Light Estimate}\label{app:diffuse_light_confirm}

\begin{figure}[h]
    \centering
    \includegraphics[scale = 0.45]{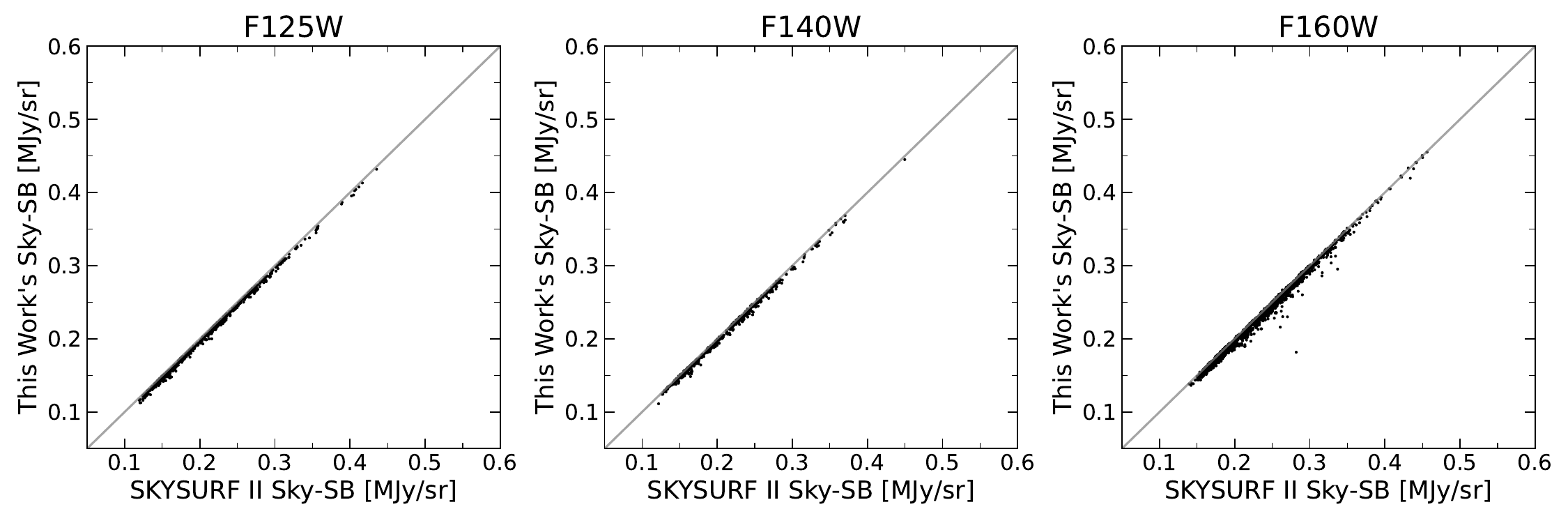}
    \caption{One-to-one relationship of the sky-SB measurements in this work (using the Pro-med algorithm) and the measurements in \citetalias{carleton_2022}. The x-axis shows sky-SB measurements from \citetalias{carleton_2022} and the y-axis shows sky-SB measurements from this work. The grey line is a one-to-one relationship.}
    \label{fig:sky_DL_1to1}
\end{figure}

We ensure our methods described in Section \ref{sec:diffuse_light_limits} result in a reliable Diffuse Light estimate. Figure \ref{fig:sky_DL_1to1} shows a one-to-one plot of sky-SB measurements used in \citetalias{carleton_2022} and those presented in this work. There are no noticeable trends in Figure \ref{fig:sky_DL_1to1}, meaning that at first order, taking a median difference between \citetalias{carleton_2022} and this work will result in consistent results, whether or not we utilize the darkest or the brightest sky-SB values.

\citetalias{carleton_2022} uses the Lowest Fitted Sky (LFS) method to estimate the DL, which utilizes the darkest sky-SB measurements. To ensure our calculated sky-SB difference represents the darkest sky-SB values necessary for the LFS method, we plot the difference in sky-SB (This Work $-$ \citetalias{carleton_2022}) as a function of the sky-SB measured in \citetalias{carleton_2022}. The red lines show the median difference used to estimate DL in Table \ref{tab:diffuse_light}. The red lines agree with the darkest sky-SB measurements, confirming that the method to estimate DL limits in this work is consistent with the LFS method from \citetalias{carleton_2022}.

\begin{figure}[h]
    \centering
    \includegraphics[scale = 0.45]{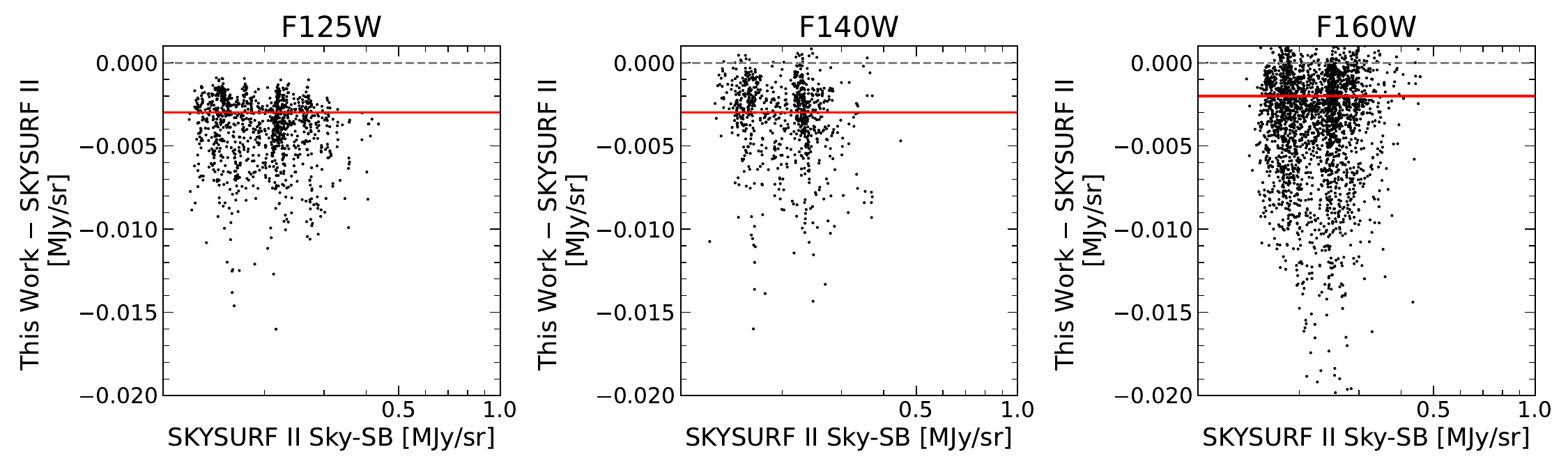}
    \caption{Relationship of the sky-SB measurements in this work (using the Pro-med algorithm) and the measurements in \citetalias{carleton_2022}. The x-axis shows sky-SB measurements from \citetalias{carleton_2022} and the y-axis shows the difference in sky-SB values between this work and \citetalias{carleton_2022}. The red lines indicate the median differences shown in Table \ref{tab:diffuse_light} used for DL estimations. The red lines agree with the darkest sky-SB measurements from \citetalias{carleton_2022}.}
    \label{fig:sky_DL}
\end{figure}

\end{document}